\documentclass[12pt]{article}
\usepackage{amsmath, amssymb,amsthm}
\usepackage{graphicx,psfrag,epsf}
\usepackage{enumerate}
\usepackage{natbib} 
\usepackage{url} 

\newcommand{\blind}{1}

\usepackage{titlesec}
\titlespacing{\section}{0pt}{0pt}{0pt}
\titlespacing{\subsection}{0pt}{0pt}{0pt}

\usepackage{authblk} 
\usepackage[colorlinks=true,linkcolor=black,citecolor=blue,urlcolor=blue]{hyperref} 
\usepackage{dsfont}
\usepackage[dvipsnames]{xcolor}

\usepackage{setspace,subcaption,multirow}
\usepackage{soul}
\usepackage{tikz}
\usepackage{pdfpages}
\usepackage{comment}
\usepackage{array}
\usepackage{hyperref}
\usepackage{float}

\usepackage{booktabs}
\usepackage{threeparttable}
\usepackage{threeparttablex}
\usepackage{array}

\usepackage{amsfonts}
\usepackage{algorithm,algpseudocode}

\newtheorem{assumption}{Assumption}
\newtheorem{theorem}{Theorem}
\newtheorem{corollary}{Corollary}

\newtheorem{remark}{Remark}%

\begin{document}

\def\spacingset#1{\renewcommand{\baselinestretch}%
{#1}\small\normalsize} \spacingset{1}



\title{\bf Designing Efficient Hybrid and Single-Arm Trials: External Control Borrowing and Sample Size Calculation}

\author[1]{Yujing Gao}
\author[2]{Xiang Zhang}
\author[1]{Shu Yang\thanks{Address for correspondence: Shu Yang, Department of Statistics, North Carolina State University, NC 27695, USA. Email: syang24@ncsu.edu}}

\affil[1]{\small Department of Statistics, North Carolina State University, Raleigh, NC 27695, USA}
\affil[2]{Medical Affairs and HTA Statistics, CSL Behring, PA 19406, USA}


\if0\blind
{
\title{\bf Designing Efficient Hybrid and Single-Arm Trials: External Control Borrowing and Sample Size Calculation}
\author{}
} \fi

\date{}
\maketitle

\begin{abstract}
External controls (ECs) from historical trials or real-world data have gained increasing attention as a way to augment hybrid and single-arm trials, especially when balanced randomization is infeasible. While most existing work has focused on post-trial inference using ECs, their role in prospective trial design remains less explored. We address this gap by focusing on the sample size determination and power analysis for an experimental design problem that encompasses standard randomized controlled trials (RCTs), hybrid trials, and single-arm trials. Building on estimators derived from the efficient influence function, we develop hybrid and single-arm design strategies that leverage comparable EC data to reduce the required sample size of the current study. We derive asymptotic variance expressions for these estimators in terms of interpretable, population-level quantities and introduce a pre-experimental variance estimation procedure to guide sample size calculation, ensuring prespecified type I error and power for the relevant hypothesis test. Simulation studies demonstrate that the proposed hybrid and single-arm designs maintain valid type I error and achieve target power across diverse scenarios while requiring substantially fewer subjects in the current study than RCT designs. A real data application further illustrates the practical utility and advantages of the proposed hybrid and single-arm designs.
\end{abstract}

\noindent%
\vfill
\newpage
\spacingset{1.9} 

\section{Introduction}
\label{sec:intro}

Randomized controlled trials (RCTs) are widely regarded as the gold standard for confirmatory clinical studies due to their ability to reduce potential confounding through randomization. However, RCTs come with drawbacks such as high costs, lengthy recruitment periods, and practical challenges, especially in rare diseases or situations with ethical constraints. Historical trial data or real-world data (RWD)  offer several advantages to complement the limitations of RCTs, including longer observation windows, access to larger and more diverse patient populations, and reduced burden on investigators and patients \citep{visvanathan2017untapped, colnet2024causal}. 
 The growing accessibility of historical data and RWD has motivated interest in leveraging external controls (ECs) to enhance trial efficiency and feasibility. ECs can be used to augment the control arm in hybrid trials or serve as comparators in single-arm trials (SATs) when randomization is infeasible due to the efficacy and safety of the investigational therapy. These approaches have gained increasing attention from industry and regulatory agencies, including the FDA, which has issued guidance outlining potential issues of using the RWD and ECs in regulatory decision making \citep{fda2021registries, FDA2023}.


Statistical methods for incorporating ECs have developed considerably since \cite{pocock1976combination}. Weighting approaches such as propensity score, empirical likelihood, entropy balancing, and constrained maximum likelihood have been proposed to address selection bias in nonrandomized studies \citep{qin2015using, chatterjee2016constrained, 
zhang2020generalized, chu2023targeted, lee2023improving, wu2023transfer, zhou2024causal}. To handle outcome distribution differences between concurrent controls and ECs, methods including test-then-pool, matching and bias adjustment, power priors, commensurate priors, and Bayesian hierarchical models \citep{ibrahim2000power, stuart2008matching, 
neuenschwander2010summarizing, Hobbs2011,
viele2014use, schoenfeld2019design, zhu2024enhancing} have been developed. However, many existing approaches rely on parametric assumptions and are vulnerable to unmeasured confounding. Estimators derived from the efficient influence function (EIF), provide a promising alternative by achieving the semiparametric efficiency bound under certain conditions while retaining double robustness \citep{li2023improving, valancius2024causal, gao2025improving, wang2025rate}.

Substantial progress has been made in developing EIF-based estimation and inference methods for trials with external control borrowing, however, their application in trial designs remains limited. Most existing hybrid design methods rely on interim analyses to guide borrowing, including sample size re-estimation using inverse probability weighting \citep{kojima2025sample}, propensity score weighting and matching methods \citep{yuan2019design,
li2022matching}, Bayesian adaptive hybrid design with power prior \citep{tian2025beam}, and adaptive hybrid design using a propensity score-based weighted estimator \citep{guo2024adaptive}. While these adaptive design methods can reduce required sample size by leveraging historical or interim data, they seldom incorporate semiparametrically efficient or doubly robust estimators, which could potentially limiting their overall efficiency and robustness. A fundamental component of trial design is sample size determination, which depends critically on the variance properties of the chosen estimator \citep{schuler2021designing}. As emphasized by \cite{tsiatis2007semiparametric}, no “reasonable” estimator can achieve an asymptotic variance lower than that of a semiparametrically efficient estimator. Therefore, incorporating ECs within an experimental design that explicitly leverages estimators possessing double robustness and semiparametric efficiency offers a promising direction to enhance trial efficiency.

In this paper, we focus on an experimental design problem that incorporates ECs without necessitating interim adaptations, while simultaneously fully leveraging EIF-based doubly robust estimators \citep{li2023improving, gao2025improving, wang2025rate}.  
We provide a generalized power formula for the consistent estimator of the target estimand, the average treatment effect (ATE) in the current study population, which is applicable to trials both with and without external control information. To enable the sample size calculation prior to trial design, we derive the asymptotic variance expression of the corresponding estimators and introduce a pre-experimental variance estimation procedure applicable to standard RCT designs, hybrid designs, and single-arm designs. By borrowing information from ECs, hybrid and single-arm designs can yield substantial sample size savings, particularly in unbalanced designs with larger treatment allocations in the current study.

The remainder of this article is organized as follows. Section \ref{sec:basic} introduces the setting, notation, and EIF-based doubly robust estimators for trials with external control borrowing. Section \ref{sec:frame} presents an experimental design problem that encompasses RCT, hybrid and single-arm designs, and provides a sufficient condition for sample size determination. Section \ref{sec:var} derives the asymptotic variance of the corresponding estimators and outlines a pre-experimental variance estimation procedure. Section \ref{sec:simu} evaluates the performance of the proposed designs through simulation studies, and Section \ref{sec:case} illustrates designs in a real-data application. Section \ref{sec:discuss} concludes with a discussion.

\section{Setting, Notation and Estimation} 
\label{sec:basic}

\subsection{Setting and notation}

We consider a one-stage, two-arm randomized controlled study comparing an active treatment to a control, evaluated at a single time point. Let $\mathcal{R}$ denote the current study and $\mathcal{E}$
the EC data, with sizes
$N_{\mathcal{R}}$ and
$N_{\mathcal{E}}$, respectively, and the total sample size is $N=N_{\mathcal{R}}+N_{\mathcal{E}}$. Within $\mathcal{R}$, $N_{t}$ subjects receive the active treatment and $N_{c}$ subjects receive the internal control, while all $N_{\mathcal{E}}$ subjects in $\mathcal{E}$ serve as the external control. For each subject $i \in \mathcal{R} \cup \mathcal{E}$, let $R_{i}$ denote the data source, where $R_i = 1$ for the current study and $R_i = 0$ for the EC, and let $A_i$ denote the treatment assignment, where $A_i = 1$ for treatment and $A_i = 0$ for control. Let $Y_i$ be the observed outcome and $\boldsymbol{X}_i$ be the baseline covariates. Under the potential outcome framework \citep{rubin1974estimating}, each subject has potential outcomes $Y_i(0)$ and $Y_i(1)$ with only one observed depending on the actual treatment received, which is formalized by the consistency assumption:
\begin{assumption}[Consistency]
\label{assum:consis}
$Y=AY(1)+(1-A)Y(0)$.
\end{assumption}

The primary causal estimand is the ATE in the current study population, defined as $\tau=\mu_{1}-\mu_{0}$, where $\mu_{a}=\mathbb{E}\{Y(a)\mid R=1\}$
for $a=0,1$. In SATs, where only the treatment group is included in the current study, the target estimand reduces to the average treatment effect on the treated (ATT). Table \ref{table_notation} summarizes the notation used in subsequent sections.

\renewcommand{\arraystretch}{1.4}  

\begin{table}[ht]
\centering
\small
\caption{\textit{Notation and description}}
\label{table_notation}
\begin{tabular}{p{6.25cm} p{9.5cm}}
\hline
\hline
\textbf{Formula}  & \textbf{Description} \\
\hline 
$\mu_{a}(\boldsymbol{X})=\mathbb{E}\{Y(a)\mid \boldsymbol{X},R=1\}$  & 
outcome mean models for $a=0,1$ in the current study, identified as $\mathbb{E}(Y\mid \boldsymbol{X},A=a,R=1)$ under Assumption \ref{assum:ign}. \\
\hline
$\pi_{A}(\boldsymbol{X})=\mathbb{P}(A=1\mid \boldsymbol{X},R=1)$  & 
treatment propensity score model in the current study, $\pi_{A}(\boldsymbol{X})=\pi_{A}$ in completely randomized study. \\ 
\hline 
$\sigma_{a,r}^{2}(\boldsymbol{X})=\mathbb{V}\{Y(a)\mid \boldsymbol{X},R=r\}$  & conditional variability of $Y(a)$ given
$\boldsymbol{X}$ in the data source $r$ for  $a=0,1$, $r=0,1$. When $r=1$, this is identified as $\mathbb{V}(Y\mid \boldsymbol{X},A=a,R=1)$ for $a=0,1$. \\
\hline 
$\sigma_{a,r}^{2}=\mathbb{V}\{Y(a)\mid R=r\}$  & marginal variability of $Y(a)$ in the data source $r$ for $a=0,1$, $r=0,1$. When $r=1$, this is identified as $\mathbb{V}(Y\mid A=a, R=1)$ for $a=0,1$.\\ 
\hline
$\pi_R(\boldsymbol{X}) = \mathbb{P}(R=1 \mid \boldsymbol{X})$ & selection propensity score model.  \\
\hline
$d(\boldsymbol{X})=\frac{f(\boldsymbol{X}\mid R=1)}{f(\boldsymbol{X}\mid R=0)}$  & ratio of the conditional density of $\boldsymbol{X}$ in the current study versus the EC.\\
\hline
$r_{R}=\frac{\mathbb{P}(R=1)}{\mathbb{P}(R=0)}$ & marginal sampling odds, with approximated
by $\frac{N_{\mathcal{R}}}{N_{\mathcal{E}}}$.\\
\hline 
$q(\boldsymbol{X})=\frac{\pi_R(\boldsymbol{X})}{ 1 - \pi_R(\boldsymbol{X})}$  & selection propensity density ratio, $q(\boldsymbol{X})$ can also be written as 
$d(\boldsymbol{X})r_{R}$. \\ 
\hline
$r(\boldsymbol{X})= \frac{\sigma_{0,1}^{2}(\boldsymbol{X})}{\sigma_{0,0}^{2}(\boldsymbol{X})}$ & conditional variance ratio of control outcomes in the current study and the EC. \\
\hline
$\kappa_{a}^{2}=\mathbb{E}\{\sigma_{a,1}^{2}(\boldsymbol{X})\mid R=1\}$  & average variability of treated/control outcomes
$(a=1/0)$ in the current study. \\
\hline
$\gamma=Corr\{\mu_{1}(\boldsymbol{X}),\mu_{0}(\boldsymbol{X}) \mid R=1\}$  & correlation of two outcome means in the current study. \\
\hline
\hline
\end{tabular}
\end{table}

\subsection{Estimation for trials with external controls}

We first consider hybrid trials with ECs, and present the assumptions required for the identification of $\tau$ \citep{li2023improving, gao2025improving}.

\begin{assumption}[Randomization and positivity]\label{assum:ign}
(i) $Y(a)\perp\!\!\!\perp A\mid(\boldsymbol{X},R=1)$ for $a=0,1,$ and (ii) 
$0<\mathbb{P}(A=1\mid \boldsymbol{X}=\boldsymbol{x},R=1)<1$ for all $\boldsymbol{x}$ such that $ \mathbb{P} (\boldsymbol{X}=\boldsymbol{x}\mid R=1)>0$. 
\end{assumption}

\begin{assumption}[External control compatibility]
\label{assum:exchange_delta}
(i) $\mathbb{E}\left\{ Y(0)\mid \boldsymbol{X}=\boldsymbol{x}, R=0\right\} = \mathbb{E}\{Y(0)\mid \boldsymbol{X}=\boldsymbol{x},R=1\}$,
and (ii) $ \mathbb{P} (R=1\mid \boldsymbol{X}=\boldsymbol{x})>0$ for all $\boldsymbol{x}$ such that $ \mathbb{P} (\boldsymbol{X}=\boldsymbol{x},R=0)>0$.
\end{assumption}

Assumption \ref{assum:ign} reflects the standard conditions of conditional exchangeability and positivity within RCTs \citep{rosenbaum1983central, imbens2004nonparametric}. Under Assumption \ref{assum:ign}, the causal estimand $\tau$ is identifiable from the current study data alone. Assumption \ref{assum:exchange_delta} requires that the conditional expectation of
$Y(0)$ be the same in the trial data and ECs. We refer to ECs satisfying Assumption \ref{assum:exchange_delta} (i) as comparable ECs. Since including non-comparable ECs can introduce bias and undermine Assumption \ref{assum:exchange_delta}, we assume throughout that all ECs used are comparable unless noted otherwise. Under Assumptions \ref{assum:consis}, \ref{assum:ign}
and \ref{assum:exchange_delta}, $\tau$ is identifiable for hybrid trials with ECs, and the corresponding EIF-based estimator is denoted as $\widehat{\tau}_{ec}$ \citep{li2023improving, gao2025improving},
\begin{align}
\widehat{\tau}_{ec} = & \ 
\frac{1}{N_{\mathcal{R}}}\sum_{i\in\mathcal{R}\cup\mathcal{E}}R_{i}\left\{ \widehat{\mu}_{1}(\boldsymbol{X}_{i})-\widehat{\mu}_{0}(\boldsymbol{X}_{i})+\frac{A_{i}\widehat{\epsilon}_{1,i}}{\widehat{\pi}_{A}(\boldsymbol{X}_{i})}\right\}\nonumber \\
& \quad -\frac{1}{N_{\mathcal{R}}}\sum_{i\in\mathcal{R}\cup\mathcal{E}}\frac{\{R_{i}(1-A_{i})+(1-R_{i})\widehat{r}(\boldsymbol{X}_{i})\}\widehat{q}(\boldsymbol{X}_i)\widehat{\epsilon}_{0,i}}{\widehat{q}(\boldsymbol{X}_i)\{1-\widehat{\pi}_{A}(\boldsymbol{X}_{i})\}+\widehat{r}(\boldsymbol{X}_{i})},
\label{eq:ACW}
\end{align}
with nuisance estimates $\widehat{\mu}_{a}(\boldsymbol{X})$, $\widehat{\pi}_A(\boldsymbol{X})$, $\widehat{r}(\boldsymbol{X})$, $\widehat{q}(\boldsymbol{X})$ and residuals $\widehat{\epsilon}_{a,i} = Y_i - \widehat{\mu}_a(\boldsymbol{X}_i)$ for $a=0,1$.


As emphasized by \cite{li2023improving, gao2025improving}, the estimator $\widehat{\tau}_{ec}$ is doubly robust, remaining consistent for $\tau$ if either (i) the outcome mean models $\mu_a(\boldsymbol{X})$ for $a\in\{0,1\}$, or (ii) propensity score models 
$\pi_A(\boldsymbol{X})$ and
$q(\boldsymbol{X})$ are correctly specified. In addition, $\widehat{\tau}_{ec}$ is locally efficient, in the sense that its asymptotic variance attains the semiparametric efficiency bound when all working models are correctly specified. Double robustness enhances reliability by providing protection against certain forms of model misspecification, whereas local efficiency ensures narrower confidence intervals relative to other estimators and improves statistical power while maintaining type I error control in trial designs. The consistency of $\widehat{\tau}_{ec}$ does not depend on the correct specification of $r(\boldsymbol{X})$, although its efficiency does. Suppose $\widehat{r}(\boldsymbol{X})$ is consistent for $r^*(\boldsymbol{X})$, under regularity conditions for estimating nuisance functions \citep{gao2025improving}, $\widehat{\tau}_{ec}$ is asymptotically normal, $\sqrt{N_{\mathcal{R}}}(\widehat{\tau}_{ec} - \tau) \stackrel{\text{d}}{\rightarrow} \mathcal{N}(0, V_{\tau_{ec}})$, where $V_{\tau_{ec}} = \mathbb{P}(R=1)\mathbb{E}\{\psi^2_{\tau_{ec}}(\boldsymbol{X},R,A,Y; \mu_1,\mu_0,\pi_A,q,r^*)\}$ is the asymptotic variance and $\psi_{\tau_{ec}}(\boldsymbol{X},R,A,Y; \mu_1,\mu_0,\pi_A,q,r)$ denotes the EIF for $\tau$ \citep{li2023improving, gao2025improving},
\begin{align}
    \psi_{\tau_{ec}}(\boldsymbol{X},R,A,Y; \mu_1,\mu_0,\pi_A,q,r) = 
    & \ 
    \frac{R}{\mathbb{P}(R=1)}\left[\left\{\mu_1(\boldsymbol{X}) -\mu_0(\boldsymbol{X}) - \tau\right\} + \frac{A\{Y-\mu_1(\boldsymbol{X})\}}{\pi_A(\boldsymbol{X})}\right]     \nonumber \\
    & \  - \frac{q(\boldsymbol{X})\left\{R(1-A) + (1-R)r(\boldsymbol{X})\right\}\{Y - \mu_0(\boldsymbol{X})\}}{\mathbb{P}(R=1)\left[q(\boldsymbol{X})\{1-\pi_A(\boldsymbol{X})\} + r(\boldsymbol{X})\right]}.
    \label{eq:eif}
\end{align}
Thus, when $r^*(\boldsymbol{X})$ coincides with the true model $r(\boldsymbol{X})$, the estimator $\widehat{\tau}_{ec}$ achieves the semiparametric efficient bound.

We next focus on single-arm trials with ECs, and present the identification assumptions for the ATT \citep{heckman1998characterizing, hirano2003efficient, wang2025rate}, 
where $R$ denotes both the data source and treatment status. 

\begin{assumption}[Positivity]
\label{assum:pos}
$0<\mathbb{P}(R=1\mid \boldsymbol{X}=\boldsymbol{x})<1$
for all $\boldsymbol{x}$ with
$\mathbb{P}(\boldsymbol{X}=\boldsymbol{x})>0$.
\end{assumption}
\begin{assumption}[Unconfoundness]
\label{assum:uncon}
$Y(r) \perp\!\!\!\perp R \mid \boldsymbol{X}$ for $r=0,1$.
\end{assumption}

Assumption \ref{assum:pos} parallels Assumption \ref{assum:exchange_delta} (ii). Following \cite{heckman1998characterizing}, Assumption \ref{assum:uncon} can be replaced by a weaker condition of mean independence, $\mathbb{E}\left\{ Y(0)\mid \boldsymbol{X}=\boldsymbol{x}, R\right\} = \mathbb{E}\{Y(0)\mid \boldsymbol{X}=\boldsymbol{x}\}$, which corresponds to Assumption \ref{assum:exchange_delta} (i) adapted to the SAT setting. When Assumptions \ref{assum:consis}, \ref{assum:pos}
and \ref{assum:uncon} hold, an EIF-based estimator for the ATT can be obtained by setting $A_i = R_i$ and $\widehat{\pi}_{A}(\boldsymbol{X}) = 1$ in equation \eqref{eq:ACW} \citep{wang2025rate}, which is denoted as $\widehat{\tau}_{sa}$. As a special case of $\widehat{\tau}_{ec}$, $\widehat{\tau}_{sa}$ inherits the properties of double robustness and local efficiency. Under regularity conditions for estimating nuisance functions \citep{wang2025rate}, $\widehat{\tau}_{sa}$ is also asymptotically normal: $\sqrt{N_{\mathcal{R}}}(\widehat{\tau}_{sa} - \tau) \stackrel{\text{d}}{\rightarrow} \mathcal{N}(0, V_{\tau_{sa}})$, where $V_{\tau_{sa}} = \mathbb{P}(R=1)\mathbb{E}\{\psi^2_{\tau_{sa}}(\boldsymbol{X},R,Y; \mu_1,\mu_0,q)\}$ is the asymptotic variance attaining the semiparametric efficiency bound with correctly specified nuisance functions, and $\psi_{\tau_{sa}}(\boldsymbol{X},R,Y; \mu_1,\mu_0,q)$ is the EIF for $\tau$ obtained by setting $A=R$ and $\pi_{A}(\boldsymbol{X}) = 1$ in equation \eqref{eq:eif}.

\section{An Experimental Design Problem}
\label{sec:frame}

We have previously discussed the identification assumptions and estimation strategies for hybrid and single-arm trials with ECs, which primarily address inference after data collection. We now shift our focus to an experimental design problem with the hypothesis test of interest as 
\begin{align*}
H_0: \tau = 0\ \mbox{and} \ H_a: \tau\neq 0.
\end{align*}
Suppose that we have a consistent estimator $\widehat{\tau}$ for $\tau$ with the asymptotic normal distribution $\sqrt{N_{\mathcal{R}}} (\widehat{\tau} - \tau)\stackrel{\text{d}}{\rightarrow}  \mathcal{N}(0, V_{\tau})$, for fixed sample size $N_{\mathcal{R}}$, we can write the approximate normal distribution of $\widehat{\tau}$ as $\widehat{\tau} \sim \mathcal{N}(\tau, V_{\tau}/ N_{\mathcal{R}})$. Thus, based on $\widehat{\tau} \sim \mathcal{N}(\tau, V_{\tau}/ N_{\mathcal{R}})$, under the fixed alternative $H_a: \tau \neq 0$ with true nonzero effect size $\tau$, the power to detect an effect of size $\tau$ at two-sided significance level $\alpha$ is 
\begin{align}
    \mbox{Power} = \Phi \left\{ \Phi^{-1} \left(\frac{\alpha}{2}\right) + \sqrt{N_{\mathcal{R}}}\frac{\tau}{\sqrt{V_{\tau}}}\right\} + \Phi\left\{ \Phi^{-1} \left(\frac{\alpha}{2}\right) - \sqrt{N_{\mathcal{R}}}\frac{\tau}{\sqrt{V_{\tau}}}\right\},
    \label{eq:power_gen}
\end{align}
where $\Phi(\cdot)$ is the standard normal cumulative distribution function. In completely randomized designs with constant allocation ratio ($\pi_A(\boldsymbol{X}) = \pi_A$), the trial design is characterized by two parameters: the treatment allocation ratio $\pi_A$ and the current study sample size $N_{\mathcal{R}}$. Specifically, $N_{\mathcal{R}}\pi_A$ subjects are assigned to the treatment, and $N_{\mathcal{R}}(1-\pi_A)$ subjects receive the internal control. For hybrid and single-arm trials that incorporate external controls, information from the EC data does not enter the power expression directly. Instead, it enters through the asymptotic variances of the corresponding estimators, 
$V_{\tau_{ec}}$ and $V_{\tau_{sa}}$, which in turn influence the power analysis and required sample size. To achieve a power greater than $1-\beta$, we determine $\pi_A$ and $N_{\mathcal{R}}$ such that the value of equation \eqref{eq:power_gen} exceeds $1-\beta$.

Given $\pi_A$, a sufficient condition on $N_{\mathcal{R}}$ to ensure the power \eqref{eq:power_gen} greater than $1-\beta$ is 
\begin{align}
    N_{\mathcal{R}} \geq \frac{V_{\tau}}{\tau^2} \left\{\Phi^{-1}(1-\beta) - \Phi^{-1}\left(\frac{\alpha}{2}\right)\right\}^2,
    \label{equ:suff_con}
\end{align}
and a conservative minimum $N_{\mathcal{R}}$ can be obtained by solving \eqref{equ:suff_con} at the boundary equality. The proof of condition \eqref{equ:suff_con} is provided in the Web Appendix A.1 in the Supplementary Materials.

When no EC data are available, the setting reduces to a completely RCT design ($N_{\mathcal{E}} = 0$, $\pi_A(\boldsymbol{X}) = \pi_A$). In this case, the standard two-sample z-test based on the difference-in-means estimator, $\widehat{\tau}_{std} = \widehat{\mathbb{E}}(Y\mid R=1, A=1) -\widehat{\mathbb{E}}(Y \mid R=1, A=0)$, is typically employed  \citep{winer1971statistical}. Under the standard assumptions for the z-test, including that outcomes are independent and identically distributed, normally distributed with known and constant variances within each RCT treatment arm, and obtained from a random sample, the estimator $\widehat{\tau}_{std}$ follows a normal distribution: $\sqrt{N_{\mathcal{R}}}(\widehat{\tau}_{std} - \tau) \sim \mathcal{N}(0, V_{\tau_{std}})$, where $V_{\tau_{std}} = \sigma_{1,1}^2/\pi_A + \sigma_{0,1}^2/(1-\pi_A)$. For a fixed allocation ratio $\pi_A$, the minimum required trial size $N_{\mathcal{R}}^{std} = N_t^{std} + N_c^{std}$ to achieve power $1-\beta$ at a two-sided significance level $\alpha$ can be derived from the power formula  \eqref{eq:power_gen} as:
\begin{align}
    N_t^{std} = \left(\sigma_{1,1}^2 + \frac{\pi_A\sigma_{0,1}^2}{1-\pi_A}\right)\frac{1}{\tau^2} \left\{\Phi^{-1}(1-\beta) - \Phi^{-1}\left(\frac{\alpha}{2}\right)\right\}^2,\ 
    N_c^{std} = \frac{(1-\pi_A)}{\pi_A} N_t^{std},
    \label{equ:size_std}
\end{align}
where $N_t^{std}$ and $N_c^{std}$ denote the required sample sizes for the treatment and control groups in the current study, respectively. Alternatively, under Assumption \ref{assum:consis} and \ref{assum:ign}, the augmented inverse probability weighted (AIPW) estimator $\widehat{\tau}_{aipw}$ is consistent and achieves the semiparametric efficiency bound with correctly specified nuisance functions, implying that no other RCT-based estimator can achieve a smaller asymptotic variance \citep{schuler2021designing}. Thus, applying $\widehat{\tau}_{aipw}$ for the RCT design is expected to improve trial efficiency relative to $\widehat{\tau}_{std}$.

When external control data are available prior to the design, they can help minimize subject enrollment in the internal control arm. In settings where the internal control arm is retained ($0 < \pi_A < 1$), $\widehat{\tau}_{ec}$ can be employed for the hybrid design. When no internal control arm is included ($\pi_A = 1$), $\widehat{\tau}_{sa}$ can be used for the single-arm design. Notably, $\widehat{\tau}_{aipw}$ arises as a special case of $\widehat{\tau}_{ec}$ obtained by setting $\widehat{r}(\boldsymbol{X}) = 0$. In both hybrid and single-arm designs, the EC data of size $N_{\mathcal{E}}$ are assumed to be fit for purpose and to satisfy the compatibility requirement. By leveraging ECs, hybrid designs have the potential to achieve greater efficiency by reducing the required sizes of the current study compared with RCT designs, especially when the treatment allocation proportion $\pi_A$ is large.

\section{Approach}
\label{sec:var}

\subsection{Variance derivation and sample size calculation}

After assigning the allocation proportion $\pi_A$, the required sample size $N_{\mathcal{R}}$ for hybrid designs, single-arm designs, and RCT designs using the AIPW estimator can be obtained by evaluating the corresponding asymptotic variance before trial initiation and substituting it into the power formula \eqref{eq:power_gen} to achieve the desired power level. We first express the asymptotic variance of $\widehat{\tau}_{ec}$ in terms of population-level parameters that can be estimated at the design stage or have natural interpretations, as presented in the following theorem. 
\begin{theorem}\label{theo:var}
    The asymptotic variance of $\widehat{\tau}_{ec}$ is 
\begin{align}
   V_{\tau_{ec}} 
   = &  \ 
   \underbrace{\mathbb{E}\left\{\frac{\sigma_{1,1}^2(\boldsymbol{X})}{\pi_A(\boldsymbol{X})} \Big| R=1\right\} }_{term \ 1(ec)} + \underbrace{ \mathbb{E}\left[\frac{\{1-\pi_A(\boldsymbol{X})\}\sigma_{0,1}^2(\boldsymbol{X})}{[\{1-\pi_A(\boldsymbol{X})\}+r(\boldsymbol{X})/q(\boldsymbol{X})]^2}\Big| R=1\right] }_{term \ 2(ec)}\nonumber \\
   & \quad 
   + \underbrace{\left\{\left(\sigma_{1,1}^2 - \kappa_1^2\right)  +   \left( \sigma_{0,1}^2 - \kappa_0^2\right) - 2\gamma \sqrt{ \left(\sigma_{1,1}^2 - \kappa_1^2\right)\left( \sigma_{0,1}^2 - \kappa_0^2\right) } \right\}}_{term\ 3(ec)}\nonumber \\
   & \quad
   +  \underbrace{\mathbb{E}\left[\frac{\{r^2(\boldsymbol{X})/r_R\}\sigma_{0,0}^2(\boldsymbol{X})}{[\{1-\pi_A(\boldsymbol{X})\} +r(\boldsymbol{X})/q(\boldsymbol{X})]^2} \Big | R=0\right]}_{term \ 4(ec)}.
   \label{eq:var}
\end{align}
\end{theorem} 

The completed proof is provided in the Appendix \ref{appendix:proof} in the Supplementary Materials. In equation \eqref{eq:var}, terms 1-3 represent contributions from the current study data, while term 4 corresponds to the EC data. Since input parameters must be specified to estimate $V_{\tau_{ec}}$ prior to the experiment, equation \eqref{eq:var} provides flexibility in trial designs by avoiding direct assumptions on the functional form or specific values of $\mu_a(\boldsymbol{X})$. 
We also derive the asymptotic variance of $\widehat{\tau}_{sa}$ for single-arm trials with ECs in the following corollary.

\begin{corollary}
\label{coro}
The asymptotic variance of $\widehat{\tau}_{sa}$ is 
\begin{align}
   V_{\tau_{sa}} 
   = &  \  \underbrace{\kappa_1^2}_{term \ 1(sa)}\nonumber + \underbrace{ \left\{\left(\sigma_{1,1}^2 - \kappa_1^2\right)  +  \left( \sigma_{0,1}^2 - \kappa_0^2\right)  - 2\gamma\sqrt{ \left(\sigma_{1,1}^2 - \kappa_1^2\right)\left( \sigma_{0,1}^2 - \kappa_0^2\right) } \right\}}_{term\ 3(sa)}\nonumber \\
   & \quad
   +  \underbrace{\mathbb{E}\left\{\frac{q^2(\boldsymbol{X})\sigma_{0,0}^2(\boldsymbol{X}) }{r_R} \Big | R=0\right\}}_{term \ 4(sa)}.
   \label{eq:sa_var}
\end{align}
\end{corollary}

Since $\widehat{\tau}_{sa}$ is a special case of $\widehat{\tau}_{ec}$, Corollary \ref{coro} follows directly by setting $\pi_A(\boldsymbol{X}) = 1$ in equation  \eqref{eq:var}.  Term 2 in $V_{\tau_{sa}}$ disappears since there is no internal control in the current study, and the remaining terms reduce to the simplified single-arm trial form in Theorem \ref{theo:var}.

Additionally, by setting $r(\boldsymbol{X})= 0$ and $\pi_A(\boldsymbol{X})=\pi_A$ in equation \eqref{eq:var}, we recover the asymptotic variance $V_{\tau_{aipw}}$ of the AIPW estimator $\widehat{\tau}_{aipw}$ in completely RCTs \citep{schuler2021designing},
\begin{align}
   V_{\tau_{aipw}}
  =  &  \left\{\sigma_{1,1}^2 + \frac{(1-\pi_A)\kappa_1^2}{\pi_A}\right\}  + \left( \sigma_{0,1}^2 + \frac{\pi_A\kappa_0^2}{1-\pi_A}\right) - 2 \gamma \sqrt{ \left(\sigma_{1,1}^2 - \kappa_1^2\right)\left( \sigma_{0,1}^2 - \kappa_0^2\right) }.
   \label{eq:var_aipw}
\end{align}
Furthermore, when the outcome mean functions are constant, $\mu_a(\boldsymbol{X}) = \mu_a$ for $a=0,1$, we have $\kappa_a^2 =\sigma_{a,1}^2$ and $V_{\tau_{aipw}}$ degenerate to $V_{\tau_{std}}$. In this case, the variance of the AIPW estimator coincides with that of the difference-in-means estimator $\widehat{\tau}_{std}$.

Given the asymptotic variance expression in equation \eqref{eq:var} and the sufficient condition in equation \eqref{equ:suff_con}, we can determine whether an sample size $N_{\mathcal{R}}$ exists that achieves the target power while controlling the type I error. For hybrid designs ($0<\pi_A<1$, $N_{\mathcal{E}} > 0$), the boundary equality in \eqref{equ:suff_con} admits a positive solution in $N_{\mathcal{R}}$, indicating that a feasible hybrid design always exists. For single-arm designs ($\pi_A = 1$, $N_{\mathcal{E}} > 0$), a sufficient feasibility condition for the existence of 
$N_{\mathcal{R}}$ is 
$N_{\mathcal{E}} > \frac{1}{\tau^2}\left\{\Phi^{-1}(1-\beta) - \Phi^{-1}\left(\frac{\alpha}{2}\right)\right\}^2\mathbb{E}\left\{d^2(\boldsymbol{X})\sigma_{0,0}^2(\boldsymbol{X}) \big | R=0\right\}$. This feasibility condition can be used to pre-check whether a single-arm trial design can be applied. For completely RCT designs using $\widehat{\tau}_{std}$ or $\widehat{\tau}_{aipw}$, the corresponding closed-form solution for $N_{\mathcal{R}}$ follows by substituting $V_{\tau_{std}}$ or $V_{\tau_{aipw}}$ into the boundary equality of \eqref{equ:suff_con}. The proofs for the existence of $N_{\mathcal{R}}$ under each design are provided in the Appendix \ref{appendix:existence} in the Supplementary Materials.

\subsection{Pre-experimental variance estimation \label{pro_est}}

\begin{figure}[H]
\centering
\includegraphics[width=0.9\linewidth]{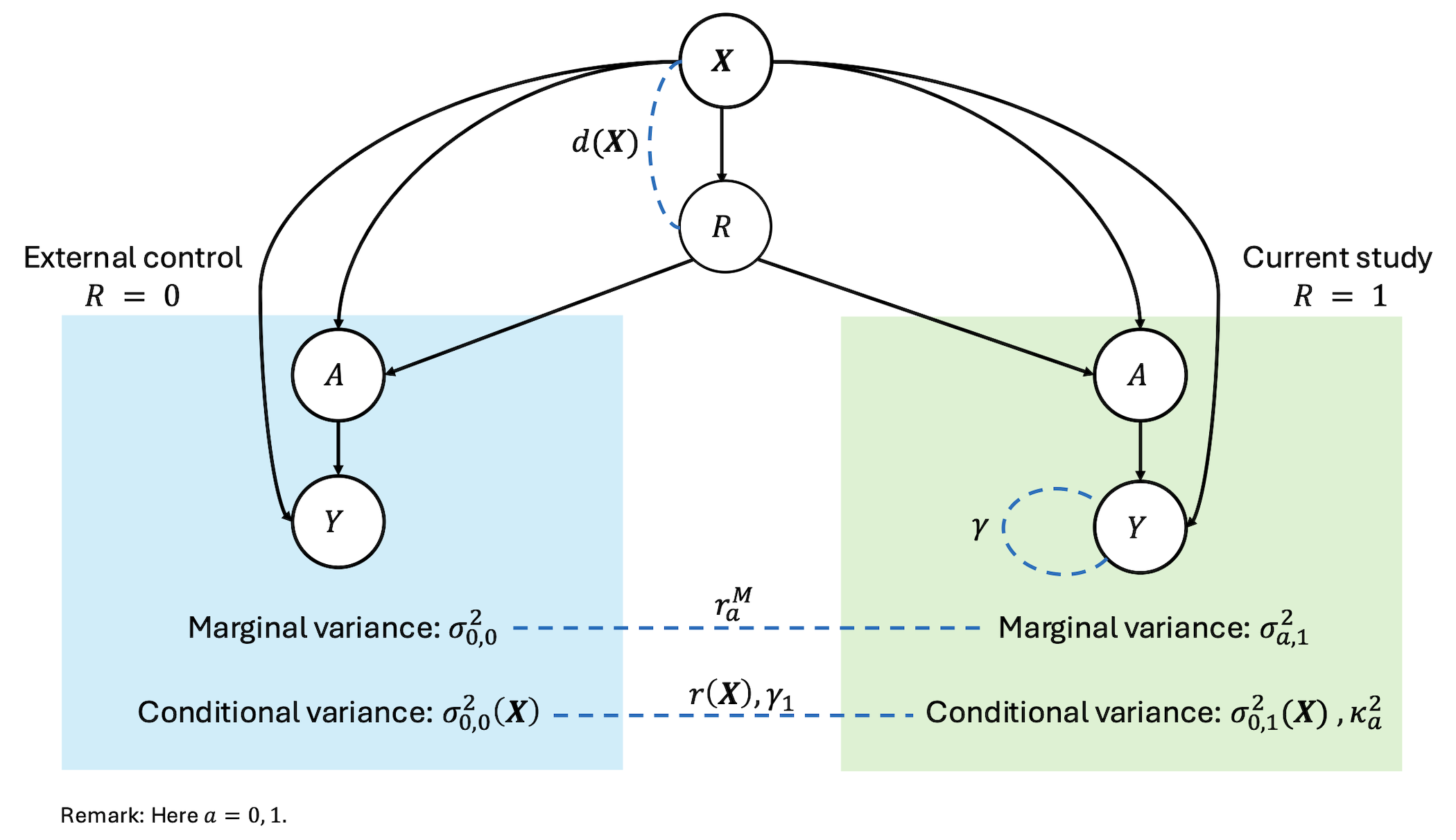}
\caption{The schematic plot for the pre-experimental variance calculation of the hybrid design, where the dashed line represents the role of the input parameter in the pre-experimental variance estimation procedure.}
\label{fig:algorithm}
\end{figure}

As shown in Figure \ref{fig:algorithm}, in a completely randomized setting with $\pi_A(\boldsymbol{X}) = \pi_A$, key components in equation \eqref{eq:var} include $\sigma_{0,1}^2(\boldsymbol{X})$, $\kappa_a^2$, $\sigma_{a,1}^2$, $q(\boldsymbol{X})$ (relying on $d(\boldsymbol{X})$), $r(\boldsymbol{X})$ and $\gamma$ for $a=0,1$, depend on the current study and cannot be identified solely from ECs prior to trial initiation. To proceed, we must either specify plausible values or bounds for these terms, or approximate them using available external control or other historical data.

To operationalize this step, we propose a pre-experimental variance estimation procedure that combines EC-based estimates with reasonable input parameters reflecting anticipated characteristics of the current study, yielding a practicable approximation to $V_{\tau_{ec}}$ for sample size calculation.


\begin{algorithm}[H]
\setlength{\baselineskip}{0.9\baselineskip}
\caption{Pre-experimental variance estimation process for design with ECs}
\label{alg1}
\begin{itemize}
    \item Step 1: Estimate $\sigma_{0,0}^2(\boldsymbol{X})$ and $\sigma_{0,0}^2$ from ECs. [default: $\widehat{\sigma}_{0,0}^2 = \frac{1}{N_\mathcal{E} -1}\sum_{i\in \mathcal{E}} \big(Y_i - \frac{1}{N_\mathcal{E}} \sum_{i\in \mathcal{E}} Y_i\big)^2$, $\widehat{\sigma}_{0,0}^2(\boldsymbol{X}) = \frac{1}{N_\mathcal{E}}\sum_{i\in \mathcal{E}} \big\{Y_i - \widehat{\mu}_0(\boldsymbol{X}_i)\big\}^2$]
    \item Step 2: Specify two positive input parameters $r_0^M$ and $r_1^M$, set $\sigma_{0,1}^2 = r_0^M\sigma_{0,0}^2$ and $\sigma_{1,1}^2 = r_1^M\sigma_{0,0}^2$. [default: $r_0^M = r_1^M = 1$] 
     \item Step 3: Specify a positive input function $r(\boldsymbol{X})$, and set $\sigma_{0,1}^2(\boldsymbol{X}) = r(\boldsymbol{X}) \sigma_{0,0}^2(\boldsymbol{X})$. [default: $r(\boldsymbol{X}) = 1$]
     \item Step 4: Specify a positive input parameter $\gamma_1$, set $\kappa_0^2 = \mathbb{E}\left\{r(\boldsymbol{X}) \sigma_{0,0}^2(\boldsymbol{X}) | R=1\right\}$ and $\kappa_1^2 = \gamma_1\kappa_0^2$. [default: $\gamma_1=1$]
    \item Step 5: If there is sufficient comparable external control data, set $d(\boldsymbol{X}) = 1$; otherwise, estimate $d(\boldsymbol{X})$ from the historical data or set it to some specific form. [default: $d(\boldsymbol{X}) = 1$]
    \item Step 6: Specify the input parameter $\gamma \in [-1,1]$. The default setting is $\gamma =1$, which corresponds to a constant additive treatment effect. [default: $\gamma =1$]
    \item Step 7: Use the above values and functions to compute $\widehat{V}_{\tau_{ec}}$ and substitute it into the power formula. 
\end{itemize}
\end{algorithm}

\begin{remark}
\label{rmk:algorithm}
    When implementing Algorithm \ref{alg1}, practitioners should first evaluate whether the available EC data are sufficiently comparable to the forthcoming study data. They should then conduct sensitivity analyses on the input parameters to assess their impact on the required sample size, based on these findings, select a conservative value for the trial design.
\end{remark}


In Step 1, we estimate the marginal variance of $Y$ and the conditional variance of $Y$ given $\boldsymbol{X}$ in the EC data. Nonparametric approaches can be used to estimate $\sigma_{0,0}^2(\boldsymbol{X})$ \citep{fan1998efficient, shen2020optimal}. We consider two simple yet practical options:  
(1) using $\widehat{\sigma}_{0,0}^2$ as a conservative upper bound, based on the law of total variance $\sigma_{0,0}^2 = \mathbb{E}\{\sigma_{0,0}^2(\boldsymbol{X})|R=0\} +  \mathbb{V}\{\mathbb{E}(Y|\boldsymbol{X},R=0)|R=0\}$; or (ii) assuming a constant conditional variance and estimating it by $\widehat{\sigma}_{0,0}^2(\boldsymbol{X}) = N_\mathcal{E}^{-1}\sum_{i\in \mathcal{E}} \{Y_i - \widehat{\mu}_0(\boldsymbol{X}_i)\}^2$ (default). Additionally, for binary outcomes, $\mathbb{E}(Y|\boldsymbol{X},R=0)$ can be modeled using logistic regression, and $\sigma_{0,0}^2(\boldsymbol{X})$ is then estimated as $\widehat{\mathbb{E}}(Y|\boldsymbol{X},R=0) \{1 -\widehat{\mathbb{E}}(Y|\boldsymbol{X},R=0)\}$. For count outcomes, $\mathbb{E}(Y|\boldsymbol{X},R=0)$ can be modeled using Poisson regression, and $\sigma_{0,0}^2(\boldsymbol{X})$ can then be estimated as $\widehat{\mathbb{E}}(Y|\boldsymbol{X},R=0)$.

In Step 2, we introduce two positive input parameters $r_1^{M}$ and $r_0^{M}$ to define the ratios of $\sigma_{1,1}^2$ and $\sigma_{0,1}^2$ in the current study data relative to $\sigma_{0,0}^2$ from the EC data. The default setting assumes $\sigma_{1,1}^2 = \sigma_{0,1}^2 = \sigma_{0,0}^2$, implying equal marginal variances of $Y$ across the treatment, internal control, and EC populations. In Step 3, we specify $\sigma_{0,1}^2(\boldsymbol{X})$ by setting $r(\boldsymbol{X})$. When $r(\boldsymbol{X}) > 1$, the conditional variance of $Y(0)$ given $\boldsymbol{X}$ is larger in the current study than in the EC; when $0<r(\boldsymbol{X}) < 1$, the variance is larger in the EC; and when $r(\boldsymbol{X})=1$ (default), the conditional variances are equal across the current study and EC populations. In Step 4, we set $\kappa_1^2 = \gamma_1\kappa_0^2$ with an input parameter $\gamma_1 > 0$. The default $\gamma_1=1$ assumes equal average variability of treated and control outcomes in the current study.

In Step 5, we specify a conditional density ratio model $d(\boldsymbol{X})$ to define $q(\boldsymbol{X}) = d(\boldsymbol{X})r_R$. When sufficient EC data are available so that, after the trial, we can match on the selection propensity score $\pi_R(\boldsymbol{X})$ and identify EC subjects similar to those in the current study (i.e., each current study subject has a similar counterpart in the EC data), this motivates the design stage default setting $d(\boldsymbol{X}) = 1$. If historical data or baseline summaries are available before the trial, $d(\boldsymbol{X})$ can instead be approximated by the ratio of the estimated conditional densities of $\boldsymbol{X}$ in the current study versus the EC, $\widehat{f}(\boldsymbol{X}\mid R=1)/ \widehat{f}(\boldsymbol{X}\mid R=0)$. An alternative approach assumes an exponential tilting model, $d(\boldsymbol{X}) = \exp\{l(\boldsymbol{X})\}/\mathbb{E}[\exp\{l(\boldsymbol{X})\}|R=0]$ for a pre-specified function $l(\boldsymbol{X})$. 
In Step 6, the correlation $\gamma$ between the two outcome means in the current study cannot be estimated solely from ECs and is therefore treated as an input parameter. Interpretable domain assumptions can be used to bound $\gamma$ \citep{schuler2021designing}. A common simplifying assumption is a constant additive treatment effect across the current study population, $\mu_1(\boldsymbol{X}) = \mu_0(\boldsymbol{X}) + \tau$, which implies $\gamma = 1$. Although this assumption may not hold exactly, we recommend assessing sensitivity to $\gamma$ to allow for treatment effect heterogeneity.

Given the specified input parameters, the pre-experimental variance estimation procedure enables the evaluation of $V_{\tau_{ec}}$ prior to trial design. The estimated variance can then be substituted into the power formula \eqref{eq:power_gen}, and the smallest $N_{\mathcal{R}}$ yielding power no less than $1-\beta$ is selected as the required sample size. When pilot data are available at the design stage, they can be used to refine assumptions and better inform input parameters. The proposed pre-experimental variance estimation procedure is applicable not only to $V_{\tau_{ec}}$ but also to its special cases, including $V_{\tau_{sa}}$ for single-arm designs, and $V_{\tau_{aipw}}$ and $V_{\tau_{std}}$ for completely RCT designs.

\section{Simulation Study}
\label{sec:simu}

\subsection{Comparison of sample sizes and assessment performance in trial designs}
\label{sec:simu_main}

We conducted simulation studies to evaluate whether the pre-experimental variance estimation results in requested sample sizes for the hybrid and single-arm designs that attain the desired power with fewer subjects than existing RCT designs using the difference-in-means estimator and the AIPW estimator. The simulation proceeded in two stages. In the first stage, we calculated the required sample size $N_{\mathcal{R}}$ for each design method. The second stage is for assessment, we evaluated the performance of each estimator by conducting the hypothesis test $H_0: \tau = 0\ vs\ H_a: \tau\neq 0$. We set the desired power level to $1-\beta=0.8$ and the significance level to $\alpha =0.05$, focusing on the unbalanced experimental design that allocates more participants to the treatment group in the current study ($0.5 \leq \pi_A < 1$). The EC data sample size was set to $N_{\mathcal{E}} = 1000$ for sufficient EC cases and $N_{\mathcal{E}} = 60$ for insufficient EC cases, across all $\pi_A$ and $N_{\mathcal{R}}$ settings.

In each scenario, the baseline covariates were $\boldsymbol{X} = (X_{1}, X_{2})$. When ECs were sufficient, $\boldsymbol{X}$ had the same distribution in the EC and current study, implying $d(\boldsymbol{X})=1$: $X_{1}^{\mathcal{R}}, X_{1}^{\mathcal{E}} \sim \mathcal{N}(1,1)$, $X_{2}^{\mathcal{R}}, X_{2}^{\mathcal{E}}\sim \mbox{Bernoulli}(p=0.5)$. For insufficient EC cases, $\boldsymbol{X}$ followed different distributions in the EC and current study, implying $d(\boldsymbol{X})\neq 1$: $X^{\mathcal{R}}_{1} \sim \mathcal{N}(1,1)$, $X^{\mathcal{E}}_{1} \sim \mathcal{N}(1.2, 1.5)$, $X^{\mathcal{R}}_{2}\sim \mbox{Bernoulli}(p=0.5)$,  $X^{\mathcal{E}}_{2}\sim \mbox{Bernoulli}(p=0.7)$. The outcome $Y$ was generated as:
\begin{align*}
   \mbox{Current study data}:& \  Y|(\boldsymbol{X},A,R =1)  = \beta_0^{\mathcal{R}} + \tau A + \beta_1^{\mathcal{R}} X^{\mathcal{R}}_{1} + \beta_2^{\mathcal{R}} X^{\mathcal{R}}_{2} +\epsilon^{\mathcal{R}}, \\
    \mbox{External control data}: & \ Y|(\boldsymbol{X},R = 0) =\beta_0^{\mathcal{E}} + \beta_1^{\mathcal{E}}X^{\mathcal{E}}_{1}+ \beta_2^{\mathcal{E}}X^{\mathcal{E}}_{2}+ \epsilon^{\mathcal{E}},
\end{align*}
where $\epsilon^{\mathcal{R}} \sim \mathcal{N}(0,0.8)$, 
$\epsilon^{\mathcal{E}} \sim \mathcal{N}(0,1)$, and the regression coefficients were set to 
$\beta_0^{\mathcal{R}} = \beta_0^{\mathcal{E}} = 1$, $\beta_1^{\mathcal{R}} = \beta_1^{\mathcal{E}} = 0.5$ and $\beta_2^{\mathcal{R}} = \beta_2^{\mathcal{E}} = -1$. The target estimand $\tau$ was constant additive across the RCT population. We set $\tau = 0$ to evaluate type I error and $\tau = 0.4$ to evaluate power.

We first determined the required sample size $N_{\mathcal{R}}$ for each design method, considering two types of sample sizes. The first type is the true required sample size, computed using the oracle values of all input parameters and nuisance functions in the pre-experimental variance estimation procedure. This represents the current study size that would achieve the target power if the data-generating mechanisms were known and reflects the sample size associated with the corresponding semiparametric efficiency bound. For the sufficient EC case, the true values were $\sigma_{0,0}^2 = 1.5$, $\sigma_{0,0}^2(\boldsymbol{X})=1$, 
$r(\boldsymbol{X}) = 0.8$, $r_0^M = r_1^M = 1.3/1.5$, $\gamma_1 = 1$,  $\gamma = 1$, and $d(\boldsymbol{X}) = 1$. For the insufficient EC case, the true values were $\sigma_{0,0}^2 = 1.585$, $\sigma_{0,0}^2(\boldsymbol{X})=1$, $r(\boldsymbol{X}) = 0.8$, $r_0^M = r_1^M = 1.3/1.585$, $\gamma_1 = 1$, $\gamma = 1$, and $d(\boldsymbol{X}) = f(X_{1}^{\mathcal{R}}) f(X_{2}^{\mathcal{R}}) /\{f(X_{1}^{\mathcal{E}}) f(X_{2}^{\mathcal{E}})\}$, where $f(\cdot)$ denotes the probability density function. We denote these true required sample sizes as $N_{\mathcal{R}}^{std, true}$, $N_{\mathcal{R}}^{aipw, true}$,  $N_{\mathcal{R}}^{ec, true}$, and 
$N_{\mathcal{R}}^{sa, true}$ corresponding to the RCT design using $\widehat{\tau}_{std}$ and $\widehat{\tau}_{aipw}$, the hybrid design using $\widehat{\tau}_{ec}$, and the single-arm design using $\widehat{\tau}_{sa}$, respectively. The second type is the non-informative required sample size for the hybrid and single-arm designs, computed under default settings through the pre-experimental variance estimation procedure.
Since $\widehat{\sigma}_{0,0}^2$ and $\widehat{\sigma}_{0,0}^2(\boldsymbol{X})$ depend on the realized EC sample, we generated $2000$ EC datasets of size $N_{\mathcal{E}}$ to account for the variability due to the random EC sampling. For each replication, we recalculated the required sample size and reported the mean across 2000 replications as the final non-informative sample sizes, denoted as $N_{\mathcal{R}}^{ec}$ and $N_{\mathcal{R}}^{sa}$. The required sample size $N_{\mathcal{R}}^{std,true}$ was obtained directly from its closed-form expression \eqref{equ:size_std}. For the remaining three design methods, we employed a grid search to identify the smallest sample size $N_{\mathcal{R}}$ that yielded power greater than or equal to $0.8$. In our simulation study, since $N_{\mathcal{E}} = 60$ does not satisfy the feasibility condition of $N_{\mathcal{E}}$ in SATs, the single-arm design was evaluated only under the sufficient EC cases ($N_{\mathcal{E}}=1000$).

\begin{table}[ht]
\centering
\caption{
\textit{Required sample size for different design methods: the RCT design using the difference-in-means estimator ($N_{\mathcal{R}}^{std,true}$), the RCT design using the AIPW estimator ($N_{\mathcal{R}}^{aipw,true}$), the hybrid design ($N_{\mathcal{R}}^{ec,true}$, $N_{\mathcal{R}}^{ec}$), and the single-arm design ($N_{\mathcal{R}}^{sa,true}$, $N_{\mathcal{R}}^{sa}$). $R^{aipw,true}$, $R^{ec,true}$,  $R^{sa,true}$ represent the percentage of true required sample size saved compared to the RCT analyzed with difference-in-means estimator, using the AIPW estimator, the hybrid design, the single-arm design, respectively.}}
\label{table_size}
\begin{tabular}{cccccccccc} 
 \hline
 \hline
 \multicolumn{10}{ c }{\textbf{Sufficient EC Cases ($N_{\mathcal{E}} = 1000$)}}\\
 \hline 
   $\pi_A$&  $N_{\mathcal{R}}^{std,true}$ & $N_{\mathcal{R}}^{aipw,true}$  & $N_{\mathcal{R}}^{ec,true}$  & 
    $N_{\mathcal{R}}^{ec}$ & $N_{\mathcal{R}}^{sa, true}$ &  $N_{\mathcal{R}}^{sa}$  & $R^{aipw,true}$ & $R^{ec,true}$ & $R^{sa,true}$ \\
 \hline
0.5 & 256 & 157 & 83 & 103 & 42 & 52 & 38.67\% & 67.58\% & 83.59\% \\
0.6 & 267 & 164 & 69 & 86  & 42 & 52 & 38.58\% & 74.16\% & 84.27\%\\
0.7 & 305 & 187 & 59 & 74 & 42 & 52 & 38.69\% & 80.66\% & 86.23\%  \\
0.8 & 399 & 246 & 52 & 65 & 42 & 52 & 38.35\% & 86.97\% & 89.47\%  \\
0.9 & 709 & 437 & 46 & 58 & 42 & 52 & 38.36\% & 93.51\% & 94.08\%  \\
 \hline
 \hline
 \multicolumn{10}{ c }{\textbf{Insufficient EC Cases ($N_{\mathcal{E}} = 60$)}}\\
 \hline
   $\pi_A$&  $N_{\mathcal{R}}^{std,true}$ & $N_{\mathcal{R}}^{aipw,true}$ 
   & $N_{\mathcal{R}}^{ec,true}$ & $N_{\mathcal{R}}^{ec}$ & $N_{\mathcal{R}}^{sa, true}$  & $N_{\mathcal{R}}^{sa}$ & $R^{aipw, true}$ & $R^{ec, true}$ &  $R^{sa,true}$ \\
 \hline
0.5 & 256 & 157  & 126 & 146 & - & - & 38.67\% & 50.78\% & -\\
0.6 & 267 & 164  &  118 & 135 & - & - & 38.58\% & 55.81\% & -\\
0.7 & 305 & 187  &  116& 130 & - & - & 38.69\% & 61.97\% & -\\
0.8 & 399 & 246  & 124 & 132 & - & - & 38.35\% & 68.92\% & -\\
0.9 & 709 & 437  & 153 & 148 & - & - & 38.36\% & 78.42\% & -\\
 \hline
 \hline
\end{tabular}
\end{table}

As shown in Table \ref{table_size}, the true required sample sizes for the hybrid and single-arm designs are consistently smaller than those of RCT designs without EC borrowing across all scenarios. Relative to the RCT design using the difference-in-means estimator, the AIPW estimator reduces the sample size by approximately 38.50\% across all cases. The hybrid design achieves larger reductions, exceeding 67.58\% with sufficient ECs and 50.78\% with insufficient ECs, while the single-arm design reduces by more than $83.59\%$ with sufficient ECs. In sufficient EC settings, the required sample size for the hybrid design decreases as $\pi_A$ increases; in insufficient EC settings, it is not necessarily monotonic in $\pi_A$, although the percentage of sample size saved still increases with $\pi_A$, reaching up to 78.42\%. Additionally, there are discrepancies between $N_{\mathcal{R}}^{ec,true}$ and  $N_{\mathcal{R}}^{ec}$, and between $N_{\mathcal{R}}^{sa,true}$ and $N_{\mathcal{R}}^{sa}$. The non-informative sample sizes generally exceed their oracle counterparts, indicating that the default setting is typically more conservative. These differences arise from 
variation in input parameter settings and the accuracy of estimating $\sigma_{0,0}^2$ and $\sigma_{0,0}^2(\boldsymbol{X})$. In practice, we recommend improving the estimation of these quantities and exploring multiple input parameter settings to identify the most conservative sample size estimate, thereby ensuring the effectiveness of subsequent trials.

In the second stage, we evaluated whether the power of $\widehat{\tau}_{ec}$ (for $N_{\mathcal{R}}>N_{\mathcal{R}}^{ec,true}$) and  $\widehat{\tau}_{sa}$ (for $N_{\mathcal{R}}>N_{\mathcal{R}}^{sa,true}$) exceeded the target design power and whether the type I error rate was controlled. We also examined the extent to which $N_{\mathcal{R}}^{ec, true}$ and $N_{\mathcal{R}}^{sa, true}$ were smaller than $N_{\mathcal{R}}^{std,true}$ and $N_{\mathcal{R}}^{aipw,true}$ across different $\pi_A$ settings. The estimators paired with their design methods were: ``Hybrid"($\widehat{\tau}_{ec}$, $\widehat{V}_{\tau_{ec}}$),  ``Single-arm"($\widehat{\tau}_{sa}$, $\widehat{V}_{\tau_{sa}}$), ``AIPW"($\widehat{\tau}_{aipw}$, $\widehat{V}_{\tau_{aipw}}$) and 
``Diff-in-mean" ($\widehat{\tau}_{std}$, $\widehat{V}_{\tau_{std}}$), with the first three calculated using the influence function. For each $\pi_A$, we vary $N_{\mathcal{R}}$ from $6$ to $500$, repeating the experiment $2000$ times per setting. For each experiment, we recorded the rejection of the null hypothesis (denoted as $1$) or not (denoted as $0$). We computed power (when $\tau=0.4$) or type I error (when $\tau=0$) as the proportion of rejections across 2000 experiments. Here we present results for $\pi_A = 0.6, 0.8$ and provide additional results for other $\pi_A$ settings in the Appendix \ref{appendix:simu2} in the Supplementary Materials.

\begin{figure}[ht]
\centering
\begin{subfigure}[b]{\textwidth}
    \centering
    \includegraphics[width=\textwidth]{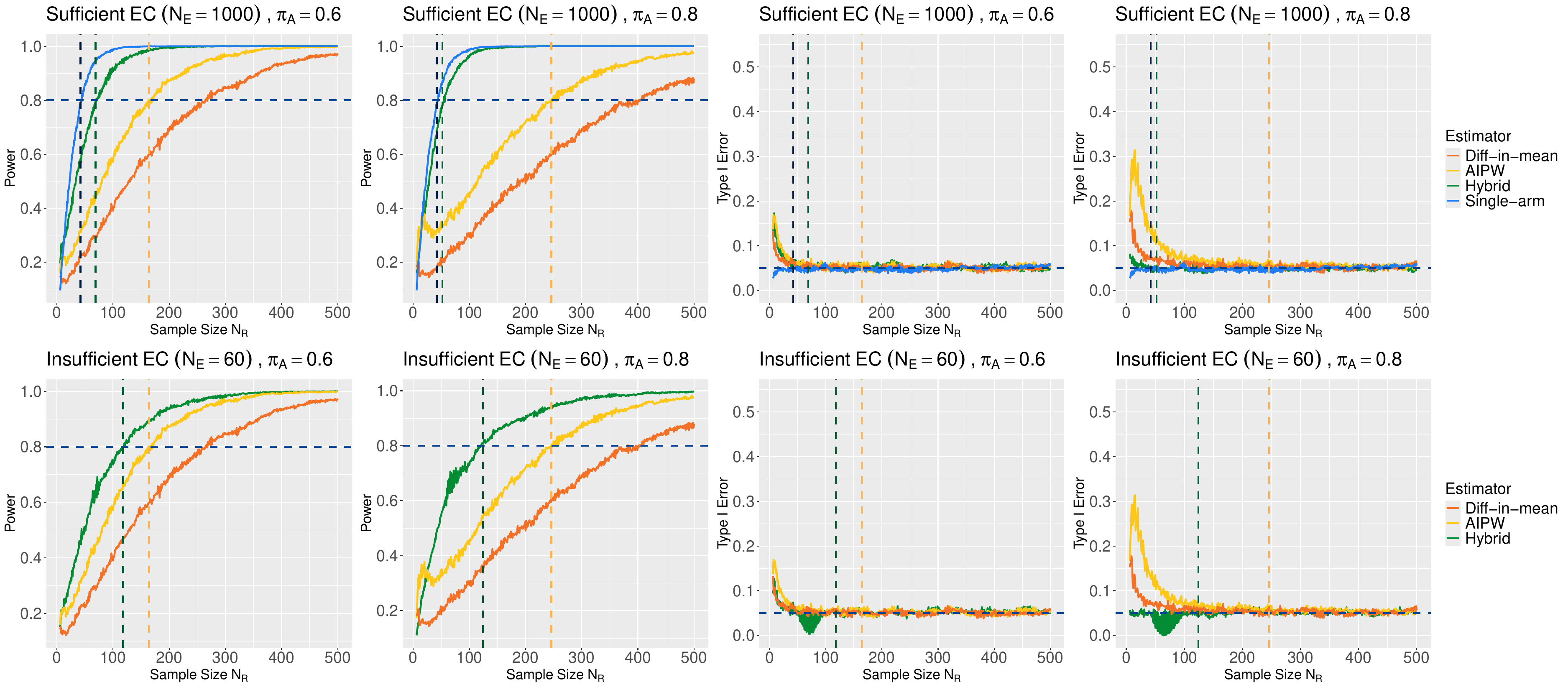}
\end{subfigure}
\vskip\baselineskip
\caption{\textit{Power and type I error of different design methods for both sufficient EC ($N_{\mathcal{E}} = 1000$) and insufficient EC ($N_{\mathcal{E}} = 60$) cases under $\pi_A = 0.6, 0.8$: the RCT design using the difference-in-means estimator, the RCT design using the AIPW estimator, the hybrid design method, and the single-arm design method. The upper panel is for sufficient EC cases and the lower panel is for insufficient EC cases. The left panel is the power plot and the right panel is the type I error plot. The horizontal dashed line represents the desired power $0.8$ in the power plot and the target type I error $0.05$ in the type I error plot. The vertical dashed lines indicate the true required sample sizes for different design methods, shown from left to right: dark blue for $N_{\mathcal{R}}^{sa,true}$, green for $N_{\mathcal{R}}^{ec,true}$, and orange for $N_{\mathcal{R}}^{aipw,true}$. The curves represent the empirical power and the type I error across varying sample size $N_{\mathcal{R}}$.}} 
\label{fig:simu}
\end{figure}

Figure \ref{fig:simu} indicates that the hybrid and single-arm designs attain the desired power and control type I error at the predetermined sample sizes $N_{\mathcal{R}}^{ec,true}$ and $N_{\mathcal{R}}^{sa,true}$. This confirms the validity of Theorem \ref{theo:var} and Corollary \ref{coro}, and the pre-experimental variance estimation procedure. When true input parameters are used in the pre-experimental variance estimation procedure at the design stage, the resulting sample size successfully delivers the desired power in the subsequent analysis. Since $N_{\mathcal{R}}^{ec,true}$ and $N_{\mathcal{R}}^{sa,true}$ are determined prior to conducting the trial, these findings provide empirical support that our proposed hybrid and single-arm designs can be effectively applied during the design phase, provided that inference is conducted using $\widehat{\tau}_{ec}$ and $\widehat{\tau}_{sa}$ as planned.

Our results further highlight the relative efficiency of these design methods. The hybrid and single-arm designs exhibit comparable performance in both power and type I error across all scenarios. Compared with $\widehat{\tau}_{std}$, the AIPW estimator $\widehat{\tau}_{aipw}$ delivers stable power improvements across $\pi_A$, while the power gains from $\widehat{\tau}_{ec}$ and 
$\widehat{\tau}_{sa}$ increase with larger $\pi_A$. This pattern arises because, in scenarios with fewer internal control subjects in the current study, 
$\widehat{\tau}_{ec}$ and $\widehat{\tau}_{sa}$ leverage EC information, whereas $\widehat{\tau}_{aipw}$ and $\widehat{\tau}_{std}$ rely solely on internal controls in the current study and thus require larger sample sizes to achieve comparable power. Additionally, $\widehat{\tau}_{ec}$ and $\widehat{\tau}_{sa}$ consistently maintain type I error rates below $0.05$ even when $N_{\mathcal{R}}$ is small, across all $\pi_A$ settings. By contrast, the type I error rates of $\widehat{\tau}_{aipw}$ and $\widehat{\tau}_{std}$ fluctuate when $N_{\mathcal{R}}$ is small, with instability worsening as $\pi_A$ increases.

Finally, we conducted an additional simulation with more limited EC data ($N_{\mathcal{E}}=30$) to illustrate the case of extremely insufficient external controls. Since its performance was similar to the $N_{\mathcal{E}} = 60$ case, we present those results in the Appendix \ref{appendix:dx2} in the Supplementary Materials for reference.

\subsection{Sensitivity analysis for pre-experimental variance estimation procedure}
\label{sec:simu_sensi}

We performed a sensitivity analysis to assess the robustness of the pre-experimental variance estimation procedure for the hybrid and single-arm designs under sufficient EC data availability. Simulations were conducted under two scenarios: (i) heterogeneous treatment effects across the current study population, and (ii) non-constant conditional variance of the outcome across both the current study and EC populations. Detailed simulation settings and results are provided in Appendix \ref{appendix:sensi} in the Supplementary Materials.

Across the sensitivity analysis, the non-informative sample sizes exceeded the true required sample size, indicating that the default settings in the pre-experimental variance estimation procedure provide a conservative design strategy in these cases. Both the hybrid and single-arm designs effectively controlled type I error and achieved the target power when using the true required sample size, confirming the robustness of the procedure. In terms of efficiency, the hybrid and single-arm designs consistently outperformed the RCT designs using either the difference-in-means or the AIPW estimators, requiring fewer participants while maintaining statistical validity.

\section{Case Study}
\label{sec:case}

We reanalyzed data from an existing trial to illustrate how the proposed hybrid and single-arm design methods can be incorporated when prospectively powering trials. The objective was to evaluate whether the inference about the treatment effect remains consistent across these designs, despite differences in required sample sizes.

The trial we used was designed to evaluate the effect of an antidepressant drug on the continuous scores of the 17-item Hamilton Depression Rating Scale (HAMD-17) \citep{mallinckrodt2014recent, liu2024multiply, gao2025doubly}. The original study, conducted under the auspices of the Drug Information Association, collected data at baseline and weeks 1, 2, 4, 6, and 8 from participants randomized to treatment and control. For our case study, we used a modified version of this dataset that removed identifiers related to marketed drugs while preserving the essential trial features. The study enrolled 172 patients, with 84 randomized to the treatment arm and 88 to the control arm. Given the presence of patient dropouts during the study, our primary interest was in estimating the ATE on the change in HAMD-17 scores at the first post-baseline visit, which contained the most complete data. Our goal was to detect a treatment effect of $-1$ with power greater than 0.8 at a two-sided significance level of $0.05$. Baseline HAMD-17 scores and gender were included as baseline covariates $\boldsymbol{X}$ in the analysis, and the treatment allocation probability $\pi_A$ was fixed at the observed trial ratio $\pi_A = 84/172$. As no external control data were available, we constructed two synthetic EC datasets by resampling from the original control arm: one of size 1000 (EC1 data) and one of size 60 (EC2 data).

Using the synthetic EC datasets, we applied five design strategies under the pre-experimental variance estimation procedure with default settings, and calculated the non-informative required sample sizes: an RCT design using the difference-in-means estimator, an RCT design using the AIPW estimator, a hybrid design with EC1 data, a hybrid design with EC2 data,  and a single-arm design with EC1 data. We also evaluated EC2 data within the procedure to check the sufficient feasibility condition for the existence of a $N_{\mathcal{R}}$ in the single-arm setting. Since the condition requires an EC data size of at least $107$ which exceeds the EC2 data size, we did not consider a single-arm design with EC2 data.

To assess whether treatment effect estimates remained consistent across different design methods, we emulated running trials side by side. For each design method, we generated bootstrap current study samples of a size equal to its non-informative required sample size from the original trial dataset to form one replicate of the current study data, paired with EC1 data or EC2 data, then applied the estimator corresponding to each design method. This procedure was repeated $2000$ times per design to account for resampling variability. For each method, we reported the mean estimated ATE in the current study population (ATT for single-arm trials), the average estimated variance across replications, and the empirical coverage rates of the 95\% Wald confidence interval for treatment effects $-1$ and 0.

\begin{table}[ht]
\centering
\small
\caption{\textit{Case study estimation results. ``Diff-in-mean" denotes RCT design using the difference-in-means estimator; ``AIPW" denotes RCT design using the AIPW estimator; ``Hybrid (EC1)" denotes hybrid design with EC1 data; ``Hybrid (EC2)" denotes the hybrid design with EC2 data; ``Single-arm (EC1)" denotes the single-arm design with EC1 data. ``Mean Est ATE" and ``Mean Est Var" are the average of the estimated ATE in the current study population (ATT for single-arm trials) and the estimated variance of estimated ATE across 2000 replications. ``Coverage ($-1$)" means the coverage rate of 95\% Wald confidence intervals for $-1$ and ``Coverage (0)" means the coverage rate of 95\% Wald confidence intervals for $0$ across 2000 replications.}}
\label{table_case_study}
\begin{tabular}{ cccccc} 
 \hline
 \hline
   \textbf{Design Method} & \textbf{$N_{\mathcal{R}}$} & \textbf{Mean Est ATE} & \textbf{Mean Est Var} & \textbf{Coverage (-1)} & \textbf{Coverage (0)}\\
 \hline
 Diff-in-mean & 458 & $-0.31$ & $0.19$ & 0.65 & 0.90 \\
 AIPW & 421 & $0.06$ & $0.20$ & 0.33 & 0.95 \\
 Hybrid (EC1) & 238 & $0.09$ & $0.25$ & 0.40 & 0.95 \\
 Hybrid (EC2) & 377 & $0.05$ & $0.20$ & 0.34 & 0.96 \\
 Single-arm (EC1) & 118 & $0.04$ & $0.24$ & 0.43 & 0.96 \\ 
 \hline
 \hline
\end{tabular}
\end{table}

From Table \ref{table_case_study}, both the hybrid and single-arm designs substantially reduce the required sample size compared with the RCT design methods. With the EC1 data, the hybrid design achieved a 48.03\% sample size reduction relative to the RCT design based on the difference-in-means estimator, whereas the single-arm design achieved a 74.24\% reduction. Although the EC2 dataset had a smaller external control sample, the hybrid design still yielded a 17.69\% reduction, outperforming the 8.08\% reduction obtained by the RCT design with the AIPW estimator. Despite the different required sample sizes across the five design methods, the estimation results were broadly consistent. The mean estimated variances were similar across all five methods, and the last four designs exhibited highly comparable performance. Although the mean estimated ATE and coverage rate at the target value $-1$ for the first design differed slightly from the others, coverage at the value 0 was comparable across all designs. Overall, these findings collectively indicate that the proposed designs yield similar inferential conclusions across most simulated trial replications. Although the true treatment effect is unknown in this case study and the original trial is not repeatable, the similarity in estimated variances supports comparable inferential performance across the designs.

\section{Discussion}
\label{sec:discuss}

We proposed hybrid and single-arm design methods that incorporate 
comparable EC data using the EIF-based doubly robust estimators \citep{li2023improving, gao2025improving, wang2025rate} to reduce the sample size required for the current study while maintaining adequate power and type I error control. We derived the asymptotic variance expressions of these EIF-based estimators and developed a pre-experimental variance estimation procedure to calculate the variance prior to trial initiation, thereby enabling the sample size calculation of the current study. Simulation studies demonstrated that both the hybrid and single-arm design methods attained the desired power and type I error control at the sample size determined by the proposed variance estimation procedure under the true input parameter settings. Furthermore, by leveraging EC data, the hybrid and single-arm designs consistently achieved higher power and required smaller sample sizes than RCT designs using either the AIPW estimator or the standard z-test, with the largest efficiency gains observed under larger treatment allocation ratios in the current study.

While our simulation studies and case study focused on continuous outcomes, the proposed hybrid and single-arm designs can be naturally extended to trials with binary or count outcomes. When the target estimand is the odds ratio for a binary outcome, the hybrid design can be adapted using the estimation approach of \cite{liu2025robust}. For small-sample settings, the proposed hybrid designs can be further developed within existing hybrid trial frameworks \citep{zhu2024enhancing, liu2025robust}. Future work may also extend our design methods to survival outcomes by incorporating the EIF-based estimator for treatment-specific survival curves proposed by \cite{gao2024doubly}. Additional extensions include applications to repeated measures in longitudinal trials \citep{zhou2024causal}, multi-level treatments \citep{yang2016propensity}, and enhanced multi-stage adaptive trial designs \citep{guo2024adaptive}.

We also provide practical guidance for specifying input parameters in the pre-experimental variance estimation procedure for hybrid and single-arm designs. When default values are used, the resulting sample size may differ from that based on the true input parameters, either upward or downward. Since it is often difficult to assess the adequacy of available EC data at the design stage, we recommend examining both the default setting and alternative functional forms for $d(\boldsymbol{X})$, informed by historical data or expert knowledge. By evaluating multiple input parameter settings and adopting the largest resulting sample size, investigators can ensure a conservative and reliable approach to planning future trials.


\section*{Acknowledgements}
This project is supported by the Food and Drug Administration (FDA) of the U.S. Department of Health and Human Services (HHS) as part of a financial assistance award, U01FD007934, totaling \$2,556,429 over three years, funded by the FDA/HHS. This work is also supported by the National Institute on Aging of the National Institutes of Health under Award Number R01AG06688, totaling \$1,565,763 over four years and the National Science Foundation under Award Number SES 2242776, totaling \$225,000 over three years. The contents are those of the authors and do not necessarily represent the official views of, nor an endorsement by, FDA/HHS, the National Institutes of Health, or the U.S. Government.

\vspace*{-8pt}



\section*{Supplementary Materials}

The Supplementary Material provides proofs and additional simulation results. 

\vspace*{-8pt}


\section*{Data Availability}

The data used for the case study in this paper, collected by \cite{mallinckrodt2014recent}, are available in the Drug Information Association Missing Data at \url{https://www.lshtm.ac.uk/research/centres-projects-groups/missing-data#dia-missing-data}. All code required for the replication of the simulations and the case study is publicly available at \url{https://github.com/yujing24/efficient_design}.


\bibliographystyle{biom}
\bibliography{refer}

\appendix

\setcounter{equation}{0}
\renewcommand{\theequation}{S\arabic{equation}}
\setcounter{table}{0}
\renewcommand{\thetable}{S\arabic{table}}
\setcounter{figure}{0}
\renewcommand{\thefigure}{S\arabic{figure}}
\setcounter{theorem}{0}
\renewcommand{\thetheorem}{S\arabic{theorem}}
\newpage

\begin{center}
{\large\bf SUPPLEMENTARY MATERIAL}
\end{center}


\section{Theory}

\subsection{Proof of the sufficient condition}
\label{appendix:condition}

\begin{proof}
Given the sufficient condition, we have 
\begin{align}
   & \ N_{\mathcal{R}} \geq \frac{V_{\tau}}{\tau^2} \left\{\Phi\left(1 - \beta\right) - \Phi^{-1} \left(\frac{\alpha}{2}\right)\right\}^2 
   \label{equ:suff_con_apx}\\
   \Leftrightarrow & \ \sqrt{N_{\mathcal{R}}}\frac{\tau}{\sqrt{V_{\tau}}} \geq \Phi^{-1}\left(1 - \beta\right) - \Phi^{-1} \left(\frac{\alpha}{2}\right) 
   \mbox{ or } 
   \sqrt{N_{\mathcal{R}}}\frac{\tau}{\sqrt{V_{\tau}}} \leq -\Phi^{-1}\left(1 - \beta\right) + \Phi^{-1} \left(\frac{\alpha}{2}\right), \nonumber 
\end{align}
where
\begin{align*}
   & \sqrt{N_{\mathcal{R}}}\frac{\tau}{\sqrt{V_{\tau}}} \geq \Phi^{-1}\left(1 - \beta\right) - \Phi^{-1} \left(\frac{\alpha}{2}\right) 
    \Leftrightarrow  
    \Phi \left\{ \Phi^{-1} \left(\frac{\alpha}{2}\right) + \sqrt{N_{\mathcal{R}}}\frac{\tau}{\sqrt{V_{\tau}}}\right\} \geq 1 - \beta, \\
   & \sqrt{N_{\mathcal{R}}}\frac{\tau}{\sqrt{V_{\tau}}} \leq -\Phi^{-1}\left(1 - \beta\right) + \Phi^{-1} \left(\frac{\alpha}{2}\right)
   \Leftrightarrow  
    \Phi \left\{ \Phi^{-1} \left(\frac{\alpha}{2}\right) - \sqrt{N_{\mathcal{R}}}\frac{\tau}{\sqrt{V_{\tau}}}\right\} \geq 1 - \beta.
\end{align*}
Thus, we have
\begin{align*}
 \mbox{Power} 
 & = \Phi \left\{ \Phi^{-1} \left(\frac{\alpha}{2}\right) + \sqrt{N_{\mathcal{R}}}\frac{\tau}{\sqrt{V_{\tau}}}\right\} + \Phi\left\{ \Phi^{-1} \left(\frac{\alpha}{2}\right) - \sqrt{N_{\mathcal{R}}}\frac{\tau}{\sqrt{V_{\tau}}}\right\}\\
 & \geq 1-\beta,
\end{align*}
which proves the condition \eqref{equ:suff_con_apx} is a sufficient condition to ensure the power greater than $1-\beta$.
\end{proof}


\subsection{Proof of Theorem 1}
\label{appendix:proof}

\begin{proof}

The efficient influence function (EIF) for the average treatment effect (ATE) in the current study population is: 
     
\begin{align*}
    \psi_{\tau_{ec}} = 
    & \  
    \frac{R}{\mathbb{P}(R=1)}\left[\left\{\mu_1(\boldsymbol{X}) -\mu_0(\boldsymbol{X}) - \tau\right\} + \frac{A\{Y-\mu_1(\boldsymbol{X})\}}{\pi_A(\boldsymbol{X})}\right] \\
    & \quad 
    - \frac{q(\boldsymbol{X})\left\{R(1-A) + (1-R)r(\boldsymbol{X})\right\}\{Y - \mu_0(\boldsymbol{X})\}}{\mathbb{P}(R=1)\left[q(\boldsymbol{X})\{1-\pi_A(\boldsymbol{X})\} + r(\boldsymbol{X})\right]}\\
    = & \ 
    \underbrace{\frac{RA\{Y-\mu_1(\boldsymbol{X})\}}{\mathbb{P}(R=1)\pi_A(\boldsymbol{X})}}_{term1: \psi_{\tau_{ec},1}} \ 
    - 
    \
    \underbrace{\frac{R(1-A)q(\boldsymbol{X})\{Y-\mu_0(\boldsymbol{X})\} }{\mathbb{P}(R=1)[q(\boldsymbol{X})\{1-\pi_A(\boldsymbol{X})\}+r(\boldsymbol{X})]}}_{term2: \psi_{\tau_{ec},2}}\\
    & \quad 
    + \  \underbrace{\frac{R\left\{\mu_1(\boldsymbol{X}) -\mu_0(\boldsymbol{X}) -\tau\right\}}{\mathbb{P}(R=1)} }_{term3: \psi_{\tau_{ec},3}}  \ - \ 
    \underbrace{\frac{(1-R)r(\boldsymbol{X})q(\boldsymbol{X})\{Y-\mu_0(\boldsymbol{X})\}}{\mathbb{P}(R=1)\left[q(\boldsymbol{X})\{1-\pi_A(\boldsymbol{X})\} +r(\boldsymbol{X})\right]} }_{term4: \psi_{\tau_{ec},4}},
\end{align*}
where $\psi_{\tau_{ec}}$ is decomposed into four components $\psi_{\tau_{ec},j}$, $j=1,2,3,4$.

Since $\mathbb{E}(\psi_{\tau_{ec}}) = 0$, the asymptotic variance of $\widehat{\tau}_{ec}$, $V_{\tau_{ec}}$, can be derived by $V_{\tau_{ec}} = \mathbb{P}(R=1) \mathbb{V}(\psi_{\tau_{ec}}^2) = \mathbb{P}(R=1) \mathbb{E}(\psi_{\tau_{ec}}^2)$, where $\mathbb{V}(B)$ denotes the variance of $B$ and $\mathbb{E}(\psi_{\tau_{ec}}^2)$ can be expressed as:
\begin{align}
   \mathbb{E}(\psi_{\tau_{ec}}^2) 
    = & \ 
    \sum_{j=1}^4
    \mathbb{E}(\psi^2_{\tau_{ec},j}) - 2\mathbb{E}(\psi_{\tau_{ec},1}\psi_{\tau_{ec},2})  + 2\mathbb{E}(\psi_{\tau_{ec},1} \psi_{\tau_{ec},3})  - 2\mathbb{E}(\psi_{\tau_{ec},1} \psi_{\tau_{ec},4}) \nonumber \\
    & \quad - 2\mathbb{E}(\psi_{\tau_{ec},2} \psi_{\tau_{ec},3}) + 2\mathbb{E}(\psi_{\tau_{ec},2}\psi_{\tau_{ec},4})  - 2\mathbb{E}(\psi_{\tau_{ec},3} \psi_{\tau_{ec},4}).
    \label{equ:Epsi}
\end{align}

\underline{Part 1: Second moment of $\psi^2_{\tau_{ec},j}$, $j = 1,2,3,4.$.}

We begin by computing the second moment of each term $\psi_{\tau_{ec},j}$, $\mathbb{E}(\psi_{\tau_{ec},j}^2)$ for $j=1,2,3,4$.


For term $\psi_{\tau_{ec},1}$,
\begin{align*}
    \mathbb{E}(\psi_{\tau_{ec},1}^2) = & \  \mathbb{E} \left[\frac{RA\{Y-\mu_1(\boldsymbol{X})\}^2}{\mathbb{P}(R =1)^2\pi_A(\boldsymbol{X})^2}\right] \\
    = & \  \frac{1}{\mathbb{P}(R =1)} \mathbb{E}\left[\frac{A\{Y-\mu_1(\boldsymbol{X})\}^2}{\pi_A(\boldsymbol{X})^2} \Big| R=1\right] \\
    = & \ \frac{1}{\mathbb{P}(R =1)} \mathbb{E}\left[\frac{1}{\pi_A(\boldsymbol{X})}\mathbb{E}\left[ \{Y(1) -\mu_1(\boldsymbol{X})\}^2 | \boldsymbol{X}, R=1, A=1\right] \Big | R=1\right]\\
    = & \ \frac{1}{\mathbb{P}(R =1)} \mathbb{E}\left[\frac{1}{\pi_A(\boldsymbol{X})}\mathbb{E}\left[ \{Y(1) -\mu_1(\boldsymbol{X})\}^2 | \boldsymbol{X}, R=1\right] \Big | R=1\right]\\
    = & \ \frac{1}{\mathbb{P}(R =1)} \mathbb{E}\left\{\frac{\sigma_{1,1}^2(\boldsymbol{X})}{\pi_A(\boldsymbol{X})} \Big| R=1\right\}.
\end{align*}


For term $\psi_{\tau_{ec},2}$,
\begin{align*}
    \mathbb{E}(\psi_{\tau_{ec},2}^2) & = \ \mathbb{E}\left[\frac{R(1-A)q^2(\boldsymbol{X})\{Y-\mu_0(\boldsymbol{X})\}^2}{\mathbb{P}(R=1)^2[q(\boldsymbol{X})\{1-\pi_A(\boldsymbol{X})\}+r(\boldsymbol{X})]^2} \right] \\
    & = \  \frac{1}{\mathbb{P}(R=1)}  \mathbb{E}\left[\frac{(1-A)q^2(\boldsymbol{X})}{[q(\boldsymbol{X})\{1-\pi_A(\boldsymbol{X})\}+r(\boldsymbol{X})]^2} \{Y-\mu_0(\boldsymbol{X})\}^2\Big| R=1\right]\\
    & = \  \frac{1}{\mathbb{P}(R=1)}  \mathbb{E}\left[\frac{\{1-\pi_A(\boldsymbol{X})\}q^2(\boldsymbol{X})}{[q(\boldsymbol{X})\{1-\pi_A(\boldsymbol{X})\}+r(\boldsymbol{X})]^2}\mathbb{E}\left[ \{Y(0)-\mu_0(\boldsymbol{X})\}^2 | \boldsymbol{X}, R=1, A=0\right]\Big| R=1\right]\\
    & = \  \frac{1}{\mathbb{P}(R=1)}  \mathbb{E}\left[\frac{\{1-\pi_A(\boldsymbol{X})\}q^2(\boldsymbol{X})}{[q(\boldsymbol{X})\{1-\pi_A(\boldsymbol{X})\}+r(\boldsymbol{X})]^2}\mathbb{E}\left[ \{Y(0) -\mu_0(\boldsymbol{X})\}^2 | \boldsymbol{X}, R=1\right]\Big| R=1\right]\\
    & = \  \frac{1}{\mathbb{P}(R=1)}  \mathbb{E}\left[\frac{\{1-\pi_A(\boldsymbol{X})\}\sigma_{0,1}^2(\boldsymbol{X})}{[\{1-\pi_A(\boldsymbol{X})\}+r(\boldsymbol{X})/q(\boldsymbol{X})]^2}\Big| R=1\right].
\end{align*}


For term $\psi_{\tau_{ec},3}$, 
\begin{align*}
    \mathbb{E}(\psi_{\tau_{ec},3}^2) 
     = & \ \mathbb{E}\left[\frac{R\left[ \left\{\mu_1(\boldsymbol{X}) - \mu_1 \right\}-\left\{\mu_0(\boldsymbol{X}) -\mu_0\right\} \right] ^2}{\mathbb{P}(R=1)^2} \right]\\
    = & \ \mathbb{E}\left[\frac{R\left\{\mu_1(\boldsymbol{X}) - \mu_1 \right\}^2}{\mathbb{P}(R=1)^2} \right] + \mathbb{E}\left[\frac{R\left\{\mu_0(\boldsymbol{X}) - \mu_0 \right\}^2}{\mathbb{P}(R=1)^2} \right]  - 2 \mathbb{E}\left[\frac{R\left\{\mu_1(\boldsymbol{X}) - \mu_1 \right\} \left\{\mu_0(\boldsymbol{X}) - \mu_0 \right\}}{\mathbb{P}(R=1)^2} \right],
\end{align*}
where 
\begin{align*}
   \mathbb{E}\left[\frac{R\{\mu_1(\boldsymbol{X})-\mu_1\}^2}{P(R =1)^2}  \right] 
   & = \  \frac{1}{\mathbb{P}(R =1)}\mathbb{E}\left[ \{\mu_1(\boldsymbol{X})-\mu_1\}^2 |R=1 \right] \\
    & = \ \frac{1}{\mathbb{P}(R =1)} \mathbb{V}\{\mu_1(\boldsymbol{X})|R=1\},\\
   \mathbb{E}\left[\frac{R\{\mu_0(\boldsymbol{X}) - \mu_0\}^2}{\mathbb{P}(R=1)^2} \right] 
    & = \ \frac{1}{\mathbb{P}(R=1)} \mathbb{E}\left[ \{\mu_0(\boldsymbol{X}) - \mu_0\}^2|R=1\right]\\
    & = \ \frac{1}{\mathbb{P}(R =1)} \mathbb{V}\{\mu_0(\boldsymbol{X})|R=1\},\\
    \mathbb{E}\left[\frac{R\left\{\mu_1(\boldsymbol{X}) - \mu_1 \right\} \left\{\mu_0(\boldsymbol{X}) - \mu_0 \right\}}{\mathbb{P}(R=1)^2} \right] 
    & = \ \frac{1}{\mathbb{P}(R=1)}  \mathbb{E}\left[\left\{\mu_1(\boldsymbol{X}) - \mu_1 \right\} \left\{\mu_0(\boldsymbol{X}) - \mu_0 \right\}\big| R=1\right] \\
    & = \ \frac{1}{\mathbb{P}(R=1)} Cov\left\{\mu_1(\boldsymbol{X}),\mu_0(\boldsymbol{X}) | R=1\right\},
\end{align*}
and $Cov(B_1,B_2)$ represents the covariance between random variables $B_1$ and $B_2$.

Thus, we have 
\begin{align*}
   \mathbb{E}(\psi_{\tau_{ec},3}^2) = & \ \frac{1}{\mathbb{P}(R =1)}  \mathbb{V}\{\mu_1(\boldsymbol{X})|R=1\} +  \frac{1}{\mathbb{P}(R =1)}  \mathbb{V}\{\mu_0(\boldsymbol{X})|R=1\} \\
    & \quad - \frac{2}{\mathbb{P}(R=1)} Cov\left\{\mu_1(\boldsymbol{X}),\mu_0(\boldsymbol{X}) | R=1\right\}.
\end{align*}


For term $\psi_{\tau_{ec},4}$,
\begin{align*}
    \mathbb{E}(\psi_{\tau_{ec},4}^2) 
    & = \ \mathbb{E}\left[  \frac{(1-R)r^2(\boldsymbol{X})q^2(\boldsymbol{X})\{Y-\mu_0(\boldsymbol{X})\}^2}{\mathbb{P}(R=1)^2[q(\boldsymbol{X})\{1-\pi_A(\boldsymbol{X})\} +r(\boldsymbol{X})]^2}  \right]\\ 
    & = \ \frac{\mathbb{P}(R=0)}{\mathbb{P}(R=1)^2} \mathbb{E}\left[ \frac{r^2(\boldsymbol{X})q^2(\boldsymbol{X}) \{Y-\mu_0(\boldsymbol{X})\}^2}{[q(\boldsymbol{X})\{1-\pi_A(\boldsymbol{X})\} +r(\boldsymbol{X})]^2} \Big | R=0 \right]\\ 
    & = \ \frac{\mathbb{P}(R=0)}{\mathbb{P}(R=1)^2} \mathbb{E}\left[\frac{r^2(\boldsymbol{X})q^2(\boldsymbol{X})}{[q(\boldsymbol{X})\{1-\pi_A(\boldsymbol{X})\} +r(\boldsymbol{X})]^2} \mathbb{E}\left[ \{Y(0) -\mu_0(\boldsymbol{X})\}^2 \big | \boldsymbol{X}, R=0 \right] \Big | R=0\right]\\ 
    & = \  \frac{\mathbb{P}(R=0)}{\mathbb{P}(R=1)^2} \mathbb{E}\left[\frac{r^2(\boldsymbol{X})\sigma_{0,0}^2(\boldsymbol{X}) }{[\{1-\pi_A(\boldsymbol{X})\} +r(\boldsymbol{X})/q(\boldsymbol{X})]^2} \Big | R=0\right],
\end{align*}
where the last equality is by Assumption 3 (i).

\underline{Part 2: Expectation of $\psi_{\tau_{ec},j_1}\psi_{\tau_{ec},j_2}$, $j_1,j_2 = 1,2,3,4.$}

Since $A(1-A)$ and $R(1-R)$ appear in $\psi_{\tau_{ec},1}\psi_{\tau_{ec},2}$, $\psi_{\tau_{ec},1}\psi_{\tau_{ec},4}$, $\psi_{\tau_{ec},2}\psi_{\tau_{ec},4}$, $\psi_{\tau_{ec},3}\psi_{\tau_{ec},4}$, expectations of these terms are all zero:
\begin{align*}
    \mathbb{E}(\psi_{\tau_{ec},1}\psi_{\tau_{ec},2}) = 0,\ 
    \mathbb{E}(\psi_{\tau_{ec},1}\psi_{\tau_{ec},4}) = 0, \ 
    \mathbb{E}(\psi_{\tau_{ec},2}\psi_{\tau_{ec},4}) = 0, \ 
    \mathbb{E}(\psi_{\tau_{ec},3}\psi_{\tau_{ec},4}) = 0.
\end{align*}
 We focus on the remaining terms: $\mathbb{E}(\psi_{\tau_{ec},1} \psi_{\tau_{ec},3})$ and $\mathbb{E}(\psi_{\tau_{ec},2}\psi_{\tau_{ec},3})$.
\begin{align*}
    \mathbb{E}(\psi_{\tau_{ec},1}\psi_{\tau_{ec},3}) = & \ 
    \mathbb{E}\left[\frac{RA\{Y-\mu_1(\boldsymbol{X})\}\left\{\mu_1(\boldsymbol{X}) -\mu_0(\boldsymbol{X}) -\tau\right\}}{\mathbb{P}(R=1)^2\pi_A(\boldsymbol{X})} \right]\\
    = & \ 
    \frac{1}{\mathbb{P}(R=1)}  \mathbb{E}\left[\frac{A\{Y-\mu_1(\boldsymbol{X})\}\left\{\mu_1(\boldsymbol{X}) -\mu_0(\boldsymbol{X}) -\tau\right\}}{\pi_A(\boldsymbol{X})} \Big| R =1 \right]\\
    = & \ 
    \frac{1}{\mathbb{P}(R=1)}  \mathbb{E}\left[\frac{\left\{\mu_1(\boldsymbol{X}) -\mu_0(\boldsymbol{X}) -\tau\right\}}{\pi_A(\boldsymbol{X})} \mathbb{E}\left[\{Y(1) -\mu_1(\boldsymbol{X})\} \big | \boldsymbol{X}, R = 1, A= 1\right]\Big| R =1 \right]\\
    = & \ 
    \frac{1}{\mathbb{P}(R=1)}  \mathbb{E}\left[\frac{\left\{\mu_1(\boldsymbol{X}) -\mu_0(\boldsymbol{X}) -\tau\right\}}{\pi_A(\boldsymbol{X})} \mathbb{E}\left[\{Y(1) -\mu_1(\boldsymbol{X})\} \big | \boldsymbol{X}, R = 1\right]\Big| R =1 \right]\\
    = & \ 0.
\end{align*}

\begin{align*}
  & \ \mathbb{E}(\psi_{\tau_{ec},2}\psi_{\tau_{ec},3}) \\
 = & \  \mathbb{E}\left[\frac{R(1-A)q(\boldsymbol{X})\{Y-\mu_0(\boldsymbol{X})\}\left\{\mu_1(\boldsymbol{X}) -\mu_0(\boldsymbol{X}) -\tau\right\} }{\mathbb{P}(R=1)^2\left[q(\boldsymbol{X})\{1-\pi_A(\boldsymbol{X})\}+r(\boldsymbol{X})\right]}\right]   \\
 = & \ 
 \frac{1}{\mathbb{P}(R=1)} \mathbb{E}\left[ \frac{(1-A)q(\boldsymbol{X})\{Y-\mu_0(\boldsymbol{X})\}\left\{\mu_1(\boldsymbol{X}) -\mu_0(\boldsymbol{X}) -\tau\right\} }{q(\boldsymbol{X})\{1-\pi_A(\boldsymbol{X})\}+r(\boldsymbol{X})}\Big | R=1 \right]   \\
 = & \ 
 \frac{1}{\mathbb{P}(R=1)} \mathbb{E} \left[\frac{q(\boldsymbol{X})\left\{\mu_1(\boldsymbol{X}) -\mu_0(\boldsymbol{X}) -\tau\right\} }{q(\boldsymbol{X})\{1-\pi_A(\boldsymbol{X})\}+r(\boldsymbol{X})}\mathbb{E} \left[\{Y(0)-\mu_0(\boldsymbol{X})\} \big | \boldsymbol{X},R=1,A=0 \right] \Big | R=1 \right] \\
  = & \ 
  \frac{1}{\mathbb{P}(R=1)} \mathbb{E} \left[\frac{q(\boldsymbol{X})\left\{\mu_1(\boldsymbol{X}) -\mu_0(\boldsymbol{X}) -\tau\right\} }{q(\boldsymbol{X})\{1-\pi_A(\boldsymbol{X})\}+r(\boldsymbol{X})}\mathbb{E} \left[\{Y(0)-\mu_0(\boldsymbol{X})\} \big | \boldsymbol{X},R=1 \right] \Big | R=1 \right] \\
  = & \ 0.
\end{align*}

\underline{Part 3: Asymptotic variance of $\widehat{\tau}_{ec}$}

Based on Part 1 and Part 2, $\mathbb{E}(\psi_{\tau_{ec}}^2)$ is
\begin{align*}
    \mathbb{E}(\psi_{\tau_{ec}}^2)
   = & \  \frac{1}{\mathbb{P}(R =1)} \mathbb{E}\left\{\frac{\sigma_{1,1}^2(\boldsymbol{X})}{\pi_A(\boldsymbol{X})} \Big| R=1\right\} \\
   &  \quad  + \frac{1}{\mathbb{P}(R=1)}  \mathbb{E}\left[\frac{\{1-\pi_A(\boldsymbol{X})\}\sigma_{0,1}^2(\boldsymbol{X})}{[\{1-\pi_A(\boldsymbol{X})\}+r(\boldsymbol{X})/q(\boldsymbol{X})]^2}\Big| R=1\right]\\
   & \quad +  \frac{1}{\mathbb{P}(R =1)}  \mathbb{V}\{\mu_1(\boldsymbol{X})|R=1\} +  \frac{1}{\mathbb{P}(R =1)}  \mathbb{V}\{\mu_0(\boldsymbol{X})|R=1\}  \\
   &  \quad -  \frac{2}{\mathbb{P}(R=1)} Cov\left\{\mu_1(\boldsymbol{X}),\mu_0(\boldsymbol{X}) | R=1\right\}\\
   & \quad +  
   \frac{\mathbb{P}(R=0)}{\mathbb{P}(R=1)^2} \mathbb{E}\left[\frac{r^2(\boldsymbol{X})\sigma_{0,0}^2(\boldsymbol{X}) }{[\{1-\pi_A(\boldsymbol{X})\} +r(\boldsymbol{X})/q(\boldsymbol{X})]^2} \Big | R=0\right],
\end{align*}
where 
\begin{align*}
    \mathbb{V}\{\mu_1(\boldsymbol{X})|R=1\}
    & = \mathbb{V}[\mathbb{E}\{Y(1)|\boldsymbol{X},R=1\}|R=1],\\
    \mathbb{V}\{\mu_0(\boldsymbol{X})|R=1\} 
    & =  \mathbb{V}[\mathbb{E}\{Y(0)|\boldsymbol{X},R=1\}|R=1].
\end{align*}

Based on the law of total variance, we have 
\begin{align*}
    \mathbb{V}[\mathbb{E}\{Y(1)|\boldsymbol{X},R=1\}|R=1] &= \mathbb{V}\{Y(1)|R=1\} - \mathbb{E}[\mathbb{V}\{Y(1)|\boldsymbol{X},R=1\}|R=1], \\
    \mathbb{V}[\mathbb{E}\{Y(0)|\boldsymbol{X},R=1\}|R=1] &= \mathbb{V}\{Y(0)|R=1\} - \mathbb{E}[\mathbb{V}\{Y(0)|\boldsymbol{X},R=1\}|R=1],
\end{align*}
then 
\begin{align*}
    \mathbb{V}\{\mu_1(\boldsymbol{X})|R=1\} & =  \mathbb{V}\{Y(1)|R=1\} - \mathbb{E}[\mathbb{V}\{Y(1)|\boldsymbol{X},R=1\}|R=1]\\
    & = \sigma_{1,1}^2 - \kappa_1^2, \\
    \mathbb{V}\{\mu_0(\boldsymbol{X})|R=1\} & =  \mathbb{V}\{Y(0)|R=1\} - \mathbb{E}[\mathbb{V}\{Y(0)|\boldsymbol{X},R=1\}|R=1]\\
    & = \sigma_{0,1}^2 - \kappa_0^2.
\end{align*}
In addition, 
\begin{align*}
   Cov\left\{\mu_1(\boldsymbol{X}),\mu_0(\boldsymbol{X}) | R=1\right\}
   & =  Corr\left\{\mu_1(\boldsymbol{X}),\mu_0(\boldsymbol{X}) | R=1\right\}\sqrt{\mathbb{V}\{\mu_1(\boldsymbol{X})|R=1\}\mathbb{V}\{\mu_0(\boldsymbol{X})|R=1\}} \\
   & = \gamma \sqrt{(\sigma_{1,1}^2 - \kappa_1^2)(\sigma_{0,1}^2 - \kappa_0^2)}.
\end{align*}

Since $r_R = \mathbb{P}(R=1)/\mathbb{P}(R=0)$, the asymptotic variance of  $\widehat{\tau}_{ec}$ is 
\begin{align*}
    V_{\tau_{ec}}
   = & \  \mathbb{P}(R=1) \mathbb{E}(\psi_{\tau_{ec}}^2) \\
    = & \  \mathbb{E}\left\{\frac{\sigma_{1,1}^2(\boldsymbol{X})}{\pi_A(\boldsymbol{X})} \Big| R=1\right\} + 
   \mathbb{E}\left[\frac{\{1-\pi_A(\boldsymbol{X})\}\sigma_{0,1}^2(\boldsymbol{X})}{[\{1-\pi_A(\boldsymbol{X})\}+r(\boldsymbol{X})/q(\boldsymbol{X})]^2}\Big| R=1\right]\\
   & \quad + \left\{(\sigma_{1,1}^2 - \kappa_1^2) + (\sigma_{0,1}^2 - \kappa_0^2) - 2\gamma\sqrt{(\sigma_{1,1}^2 - \kappa_1^2)(\sigma_{0,1}^2 - \kappa_0^2)}\right\} \\
   & \quad +  \mathbb{E}\left[\frac{\{r^2(\boldsymbol{X})/r_R\}\sigma_{0,0}^2(\boldsymbol{X}) }{[\{1-\pi_A(\boldsymbol{X})\} +r(\boldsymbol{X})/q(\boldsymbol{X})]^2} \Big | R=0\right]. 
\end{align*}

\end{proof}


\subsection{Discussion for existence of $N_{\mathcal{R}}$ in each design method \label{appendix:existence}}

Given the sufficient condition \eqref{equ:suff_con_apx}, we will discuss the existence of $N_{\mathcal{R}}$ in each design method.

\begin{proof}

\underline{Part 1: Existence of $N_{\mathcal{R}}$ in hybrid designs.}

Given $0<\pi_A <1$ and $N_{\mathcal{E}} > 0$, the expression for $V_{\tau_{ec}}$ in Theorem \ref{theo:var} simplifies to:
\begin{align}
   V_{\tau_{ec}}
  = &  \frac{\kappa_1^2}{\pi_A}
  + 
  \mathbb{E}\left[\frac{(1-\pi_A)\sigma_{0,1}^2(\boldsymbol{X})}{\{(1-\pi_A)+r(\boldsymbol{X})/d(\boldsymbol{X})/r_R\}^2}\Big| R=1\right] \nonumber \\
   & \ +  \left\{\left(\sigma_{1,1}^2 - \kappa_1^2\right) +   \left( \sigma_{0,1}^2 - \kappa_0^2\right) - 2\gamma \sqrt{ \left(\sigma_{1,1}^2 - \kappa_1^2\right)\left(\sigma_{0,1}^2 - \kappa_0^2\right)}\right\} \nonumber \\
   & \ + 
   \mathbb{E}\left[\frac{\{r^2(\boldsymbol{X})/r_R\}\sigma_{0,0}^2(\boldsymbol{X})}{\{(1-\pi_A) +r(\boldsymbol{X})/d(\boldsymbol{X})/r_R\}^2} \Big | R=0\right],
   \label{eq:var_simp}
\end{align}
where $r_{\mathcal{R}} = N_{\mathcal{R}}/ N_{\mathcal{E}}$.

Define the following notations: 
\begin{align*} 
K & = \frac{\left\{\Phi^{-1}(1-\beta) - \Phi^{-1}\left(\frac{\alpha}{2}\right)\right\}^2}{\tau^2}, \\ 
E_1(N_{\mathcal{R}}) & = \mathbb{E}\left[\frac{(1-\pi_A)\sigma_{0,1}^2(\boldsymbol{X})}{\{(1-\pi_A)+r(\boldsymbol{X})/d(\boldsymbol{X})/r_R\}^2}\Big| R=1\right], \\ 
E_2(N_{\mathcal{R}}) & = \mathbb{E}\left[\frac{\{r^2(\boldsymbol{X})/r_R\}\sigma_{0,0}^2(\boldsymbol{X})}{\{(1-\pi_A) +r(\boldsymbol{X})/d(\boldsymbol{X})/r_R\}^2} \Big | R=0\right], \\ 
V_1 & = \frac{\kappa_1^2}{\pi_A} +  
\left\{\left(\sigma_{1,1}^2 - \kappa_1^2\right) +   \left( \sigma_{0,1}^2 - \kappa_0^2\right) - 2\gamma \sqrt{ \left(\sigma_{1,1}^2 - \kappa_1^2\right)\left(\sigma_{0,1}^2 - \kappa_0^2\right)}\right\},
\end{align*}
and $V_{\tau_{ec}} = V_1 + E_1(N_{\mathcal{R}}) + E_2(N_{\mathcal{R}})$.

Plug $ V_{\tau_{ec}}$ into the sufficient condition \eqref{equ:suff_con_apx}, 
\begin{align*}
 N_{\mathcal{R}} \geq K\{V_1 + E_1(N_{\mathcal{R}}) + E_2(N_{\mathcal{R}})\},
\end{align*}
and the boundary equality is 
\begin{align*}
    N_{\mathcal{R}} = K\{V_1 + E_1(N_{\mathcal{R}}) + E_2(N_{\mathcal{R}})\}.
\end{align*}

When $N_{\mathcal{R}} \rightarrow 0$, the right side $\rightarrow - KV_1$ ($K>0$ and $V_1>0$), which is negative; when $N_{\mathcal{R}} \rightarrow +\infty$, the right side $\rightarrow +\infty$. By the intermediate value theorem, the continuity of $N_{\mathcal{R}} - K\{V + E_1(N_{\mathcal{R}}) + E_2(N_{\mathcal{R}})\}$ on $N_{\mathcal{R}} \in (0,+\infty)$ implies that there exists $N_{\mathcal{R},ec}$ making the boundary equality hold.

\underline{Part 2: Sufficient feasibility condition for existence of $N_{\mathcal{R}}$ in single-arm designs.}

Given $\pi_A = 1$, the asymptotic variance of $\widehat{\tau}_{sa}$ is:
\begin{align*}
   V_{\tau_{sa}} 
   = & \  \kappa_1^2 + \left\{\left(\sigma_{1,1}^2 - \kappa_1^2\right)  +  \left( \sigma_{0,1}^2 - \kappa_0^2\right)  - 2\gamma\sqrt{ \left(\sigma_{1,1}^2 - \kappa_1^2\right)\left( \sigma_{0,1}^2 - \kappa_0^2\right) } \right\} \\
   & \quad + r_{R}\mathbb{E}\left\{d^2(\boldsymbol{X})\sigma_{0,0}^2(\boldsymbol{X})\Big | R=0\right\}, 
\end{align*}

Define the notations below:
\begin{align*}
    E_3 & = \mathbb{E}\left\{d^2(\boldsymbol{X})\sigma_{0,0}^2(\boldsymbol{X}) \Big | R=0\right\},\\
    V_2 & = \sigma_{1,1}^2 + \left( \sigma_{0,1}^2 - \kappa_0^2\right)  - 2\gamma\sqrt{ \left(\sigma_{1,1}^2 - \kappa_1^2\right)\left( \sigma_{0,1}^2 - \kappa_0^2\right) }, 
\end{align*}
and $V_{\tau_{sa}} = V_2 + \frac{N_{\mathcal{R}}}{N_{\mathcal{E}}} E_3$. Plug $V_{\tau_{sa}}$ in the boundary equality of the sufficient condition \eqref{equ:suff_con_apx}, we can derive the closed form of $N_{\mathcal{R},sa}$ as
\begin{align*}
  N_{\mathcal{R},sa} = \frac{KV_2N_{\mathcal{E}}}{N_{\mathcal{E}} - E_3 K},
\end{align*}
To make $N_{\mathcal{R},sa} > 0$ and thus $N_{\mathcal{R},sa}$ exist, we need to ensure $N_{\mathcal{E}} - E_3 K>0$, which formalizes the sufficient feasibility condition for the existence of $N_{\mathcal{R},sa}$ in single-arm design:
\begin{align*}
   N_{\mathcal{E}} >  \frac{1}{\tau^2}\left\{\Phi^{-1}(1-\beta) - \Phi^{-1}\left(\frac{\alpha}{2}\right)\right\}^2\mathbb{E}\left\{d^2(\boldsymbol{X})\sigma_{0,0}^2(\boldsymbol{X}) \Big | R=0\right\}.
\end{align*}

\underline{Part 3: Closed-forms of  $N_{\mathcal{R}}$ in randomized controlled designs.}

Given $\pi_A$, variances of the difference-in-means estimator and the augmented inverse probability weighted (AIPW) estimator are: 
\begin{align*}
V_{\tau_{aipw}} & = \left\{\sigma_{1,1}^2 + \frac{(1-\pi_A)\kappa_1^2}{\pi_A}\right\}  + \left( \sigma_{0,1}^2 + \frac{\pi_A\kappa_0^2}{1-\pi_A}\right) - 2 \gamma \sqrt{ \left(\sigma_{1,1}^2 - \kappa_1^2\right)\left( \sigma_{0,1}^2 - \kappa_0^2\right)}, \\
V_{\tau_{std}} & = \frac{\sigma_{1,1}^2}{\pi_A} + \frac{\sigma_{0,1}^2}{1-\pi_A}.
\end{align*}

Plugging these variances into the boundary equality of the sufficient condition \eqref{equ:suff_con_apx}, we can derive closed-forms of $N_{\mathcal{R}}$ in randomized controlled trials (RCTs) designs, $N_{\mathcal{R},aipw} = K V_{\tau_{aipw}}, \ N_{\mathcal{R},std} = K V_{\tau_{std}}$, where $K = \frac{\left\{\Phi^{-1}(1-\beta) - \Phi^{-1}\left(\frac{\alpha}{2}\right)\right\}^2}{\tau^2}$.
\end{proof}


\section{Supplementary simulation study}
\label{appendix:simu}

\subsection{Supplementary simulation results for Section \ref{sec:simu_main}}
\label{appendix:simu2}

\begin{figure}[H]
\centering
\includegraphics[width=0.9\linewidth]{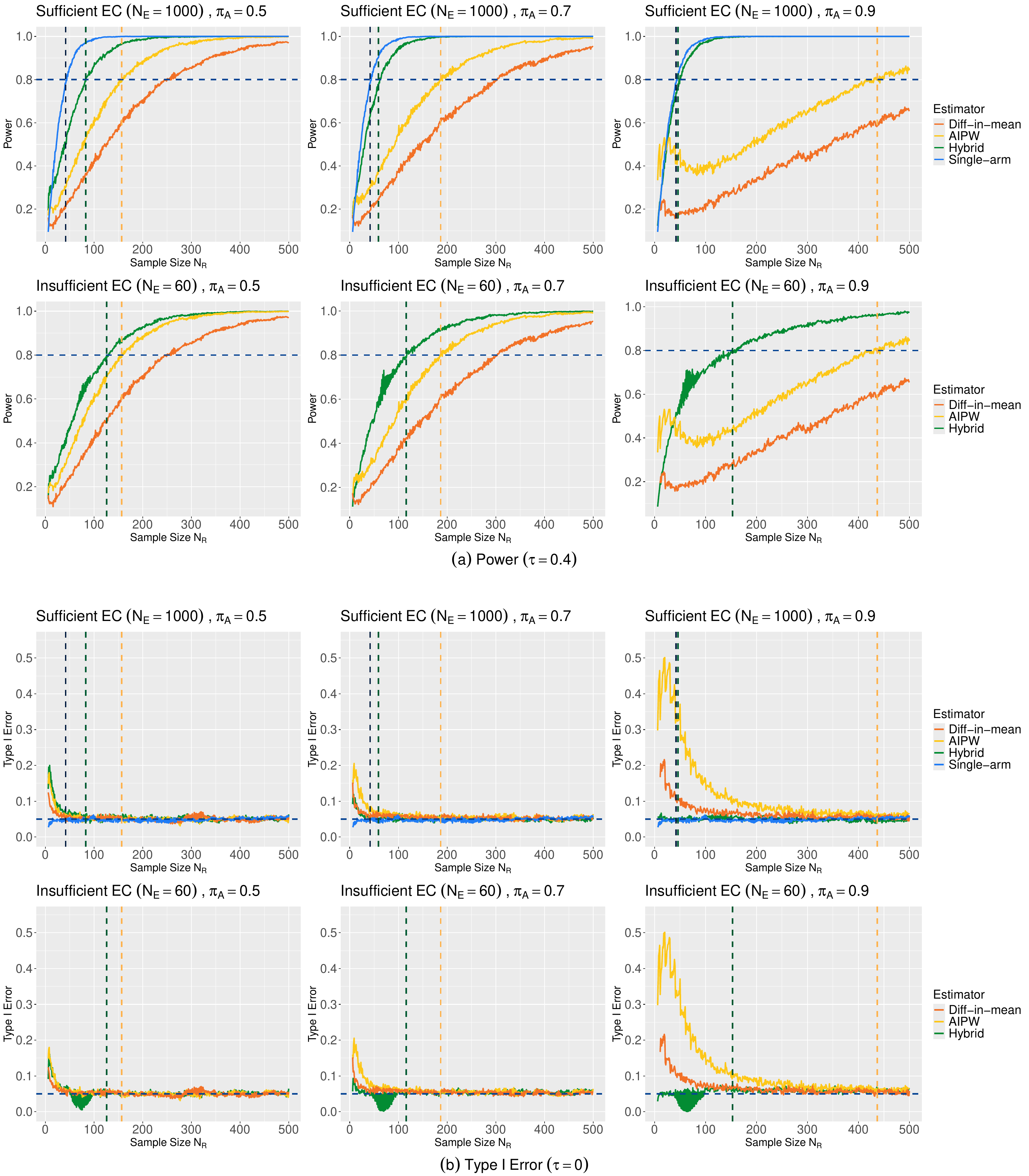}
\caption{
Power and type I error of different design methods for both sufficient EC cases ($N_{\mathcal{E}} = 1000$) and insufficient EC cases ($N_{\mathcal{E}} = 1000$) cases under $\pi_A = 0.5,0.7,0.9$: the RCT design using the difference-in-means estimator and the AIPW estimator, the hybrid design, and the single-arm design. The upper subfigure (a) is the power plot and the lower subfigure (b) is the type I error plot. In each subfigure, the upper panel is for sufficient EC cases and the lower panel is for insufficient EC cases. The horizontal dashed line represents the desired power $0.8$ in the power plot and the target type I error $0.05$ in the type I error plot. The vertical dashed lines indicate the true required sample size for each design method, shown from left to right: dark blue for $N_{\mathcal{R}}^{sa,true}$, green for $N_{\mathcal{R}}^{ec,true}$, and orange for 
$N_{\mathcal{R}}^{aipw,true}$. The curves represent the empirical power and type I error across varying sample size $N_{\mathcal{R}}$.} 
\label{appendix:fig_simu1}
\end{figure}

We provide supplementary simulation results under both sufficient external control (EC) cases ($N_{\mathcal{E}} = 1000$) and insufficient EC cases ($N_{\mathcal{E}} = 60$) in Figure \ref{appendix:fig_simu1}, with treatment assignment probabilities $\pi_A = 0.5, 0.7, 0.9$.


\subsection{Simulation results for insufficient EC cases ($N_{\mathcal{E}} = 30$)}
\label{appendix:dx2}

We kept the same simulation setting as in Section \ref{sec:simu_main} for the insufficient EC cases ($N_{\mathcal{E}} = 60$), except that the EC sample size was further reduced to $N_{\mathcal{E}} = 30$. Accordingly, the true values of the input parameters and models used for sample size calculation remain unchanged. We reported the true required sample sizes for each design method, excluding the single-arm design, ($N_{\mathcal{R}}^{std,true}$, $N_{\mathcal{R}}^{aipw,true}$, $N_{\mathcal{R}}^{ec,true}$), as well as the non-informative required sample size for the hybrid design ($N_{\mathcal{R}}^{ec}$).

\begin{table}[ht]
\centering
\caption{
\textit{Required sample sizes for the RCT design using the difference-in-means estimator ($N_{\mathcal{R}}^{std,true}$) and using the AIPW estimator ($N_{\mathcal{R}}^{aipw,true}$), and the hybrid design ($N_{\mathcal{R}}^{ec,true}$, $N_{\mathcal{R}}^{ec}$). $R^{aipw,true}$ and $R^{ec,true}$ represent the percentage of true required sample size saved compared to the RCT analyzed with the difference-in-means estimator, using the AIPW estimator, the hybrid design, respectively.}
}
\begin{tabular}{ccccccc} 
 \hline
 \hline
 \multicolumn{7}{c}{\textbf{Insufficient EC Cases ($N_{\mathcal{E}} = 30$)}}\\
 \hline 
   $\pi_A$ &  $N_{\mathcal{R}}^{std,true}$ & $N_{\mathcal{R}}^{aipw,true}$  &
   $N_{\mathcal{R}}^{ec, true}$ &
   $N_{\mathcal{R}}^{ec}$ &
   $R^{aipw,true}$ & 
   $R^{ec, true}$ \\
 \hline
0.5 & 256 & 157 & 138 & 151 & 38.67\% & 46.09\% \\
0.6 & 267 & 164  & 136 & 146 & 38.58\% & 49.06\% \\
0.7 & 305 & 187  & 143 & 152 &  38.69\% & 53.11\% \\
0.8 & 399 & 246  & 170 & 174 & 38.35\% & 57.39\%\\
0.9 & 709 & 437  & 261 & 250 &  38.36\% & 63.19\% \\
 \hline
 \hline
\end{tabular}
\label{table_size_dx2}
\end{table}

Table \ref{table_size_dx2} shows that across all values of $\pi_A$, the hybrid design consistently yields a smaller required sample size $N_{\mathcal{R}}$ than using both the difference-in-means and AIPW estimators in the RCT design. While the true required sample size for the hybrid design does not always decrease with increasing $\pi_A$, the percentage of required sample size saved increases, reaching up to 63.19\%.

Comparing the simulation results of insufficient EC cases with $N_{\mathcal{E}}= 60$ and $N_{\mathcal{E}}= 30$, we observed that for each value of $\pi_A$, the required sample size for the hybrid design increases as $N_{\mathcal{E}}$ decreases, and the percentage of sample size saved correspondingly declines. The smallest true required sample size is achieved at $\pi_A=0.7$ for $N_{\mathcal{E}}= 60$, and at 
$\pi_A = 0.6$ for $N_{\mathcal{E}}= 30$.

\begin{figure}[H]
\centering
\includegraphics[width=1\textwidth]{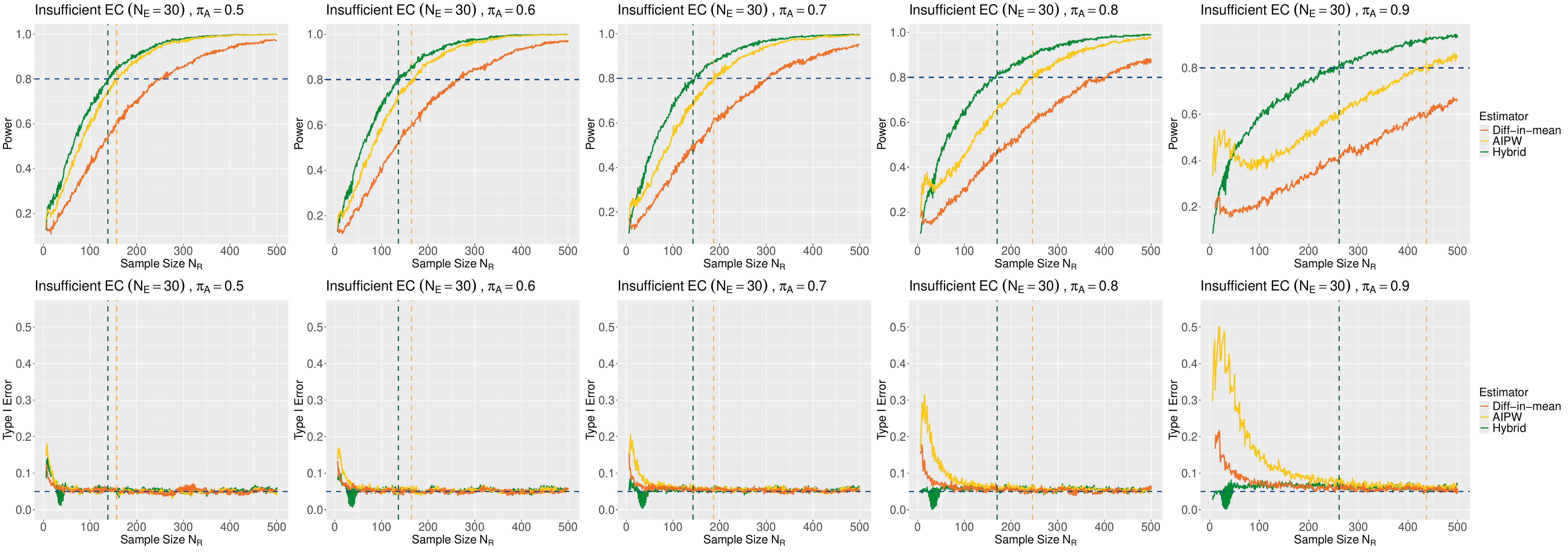}
\caption{\textit{Power and type I error of different design methods under the insufficient EC case ($N_{\mathcal{E}} = 30$) across $\pi_A = 0.5,0.6,0.7,0.8, 0.9$. Those methods include the RCT design with the difference-in-means estimator, the RCT design using the AIPW estimator, and the hybrid design method. The upper panel is the power plot and the lower panel is the type I error plot. The horizontal dashed line indicates the desired power $0.8$ in the power plot and the target type I error $0.05$ in the type I error plot. The vertical dashed lines represent the true required sample sizes for each design method: green for $N_{\mathcal{R}}^{ec,true}$ and orange for $N_{\mathcal{R}}^{aipw, true}$. The curves represent the empirical power and type I error across varying sample size $N_{\mathcal{R}}$.}
} 
\label{appendix:fig_simu2}
\end{figure}

We further evaluated the performance of the estimator in each design method by conducting the hypothesis test, $H_0: \tau = 0\ vs\ H_a: \tau\neq 0$. For each $\pi_A$, we varied $N_{\mathcal{R}}$ from $6$ to $500$, repeating the experiment $2000$ times per setting. We computed power (when $\tau=0.4$) and type I error (when $\tau=0$) as the proportion of rejections across $2000$ experiments. Simulation results in Figure \ref{appendix:fig_simu2} demonstrate that the hybrid design using $\widehat{\tau}_{ec}$ achieves power above $0.8$ and controls type I error at $0.05$ at the corresponding $N_{\mathcal{R}}^{ec,true}$. Additionally, it has a larger power than the RCT design using both the AIPW and difference-in-means estimators with the same required sample size.

\subsection{Sensitivity analysis simulation results}
\label{appendix:sensi}

We focused on simulations with sufficient EC data $(N_{\mathcal{E}} = 1000)$ and conducted a sensitivity analysis to evaluate the robustness of the proposed pre-experimental variance estimation procedure used for sample size calculation. This analysis still considered unbalanced designs requiring more treatment units in the current study data ($0.5 \leq \pi_A < 1$). Across all scenarios, we set the desired power level at $1-\beta = 0.8$ and the significance level at $\alpha = 0.05$. We reported the required sample sizes and assessed the performance of each design method based on empirical power and type I error.


\subsubsection{Simulation setting: heterogeneous treatment effects across the current study population}\label{appendix:sensi_1}

We considered the setting with a heterogeneous treatment effect across the current study population. The baseline covariates $\boldsymbol{X} = (X_{1}, X_{2})$ were generated from the same distribution in both the EC and the current study data: $X_{1} \sim \mathcal{N}(1,1)$, $X_{2}\sim \mbox{Bernoulli}(p=0.5)$. The outcome $Y$ was generated as:
\begin{align*}
   \mbox{Current study data}:& \  Y|(\boldsymbol{X},A,R =1)  = \beta_0^{\mathcal{R}} + \tau_1 A X_{1} + \tau_2AX_{2} + \beta_1^{\mathcal{R}} X_{1} + \beta_2^{\mathcal{R}} X_{2} + \epsilon^{\mathcal{R}}, \\
    \mbox{External control data}: & \ Y|(\boldsymbol{X},R = 0) =\beta_0^{\mathcal{E}} + \beta_1^{\mathcal{E}} X_{1}+ \beta_2^{\mathcal{E}} X_{2}+ \epsilon^{\mathcal{E}},
\end{align*}
where $\epsilon^{\mathcal{R}} \sim \mathcal{N}(0,0.8)$, 
$\epsilon^{\mathcal{E}} \sim \mathcal{N}(0,1)$, and the regression coefficients were set to $\beta_0^R = \beta_0^{\mathcal{E}} = 1$, $\beta_1^{\mathcal{R}} = \beta_1^{\mathcal{E}} = 0.5$, and $\beta_2^{\mathcal{R}} = \beta_2^{\mathcal{E}} = -1$. The target estimand is still constant, $\tau = \tau_1 + 0.5\tau_2$, but the treatment effect in the current study group varies with the covariate $\boldsymbol{X}$, which is not a constant additive effect. For type I error evaluation, we set $\tau_1 = \tau_2 = 0$, and for power evaluation, we set $\tau_1 = 0.3$, $\tau_2 = 0.2$, resulting in $\tau = \tau_1 + 0.5\tau_2 = 0.4$. The outcome mean model in the external controls is identical to that in the internal controls, implying this simulation setting satisfies Assumption \ref{assum:exchange_delta} (i).

True values of the input parameters and models for the pre-experimental variance estimation process were:  $\sigma_{0,0}^2 = 1.5$, $\sigma_{0,0}^2(\boldsymbol{X}) = 1$, $\kappa_0^2 = \kappa_1^2 = 0.8$, $r(\boldsymbol{X})=0.8$, $r_0^M=1.3/1.5$, $r_1^M = 1.6/1.5$, $\gamma_1 = 1$, $\gamma = 0.6/\sqrt{0.4}$, $d(\boldsymbol{X}) = 1$.


\subsubsection{Simulation setting: non-constant conditional variance of the outcome.}\label{appendix:sensi_2}

\ We next focused on the setting where the conditional variance of $Y$ given $\boldsymbol{X}$ is non-constant across both the current study and external control populations. The baseline covariates $\boldsymbol{X} = (X_{1}, X_{2})$ followed the same distribution in both the EC and the current study data: $X_{1} \sim \mathcal{N}(1,1)$, $X_{2}\sim \mbox{Bernoulli}(p=0.5)$. The outcome $Y$ was generated as:
\begin{align*}
   \mbox{Current study data}:& \  Y|(\boldsymbol{X},A,R =1)  = \beta_0^{\mathcal{R}} + \tau A + \beta_1^{\mathcal{R}} X_{1} + \beta_2^{\mathcal{R}} X_{2} + 0.8X_{1}\epsilon^{\mathcal{R}}, \\
    \mbox{External control data}: & \ Y|(\boldsymbol{X},R = 0) =\beta_0^{\mathcal{E}} + \beta_1^{\mathcal{E}} X_{1}+ \beta_2^{\mathcal{E}} X_{2}+ 0.4X_{1}^2\epsilon^{\mathcal{E}},
\end{align*}
where $\epsilon^{\mathcal{R}} \sim \mathcal{N}(0,0.8)$, 
$\epsilon^{\mathcal{E}} \sim \mathcal{N}(0,1)$, and the regression coefficients were set to $\beta_0^{\mathcal{R}} = \beta_0^{\mathcal{E}} = 1$, $\beta_1^{\mathcal{R}} = \beta_1^{\mathcal{E}} = 0.5$, and $\beta_2^{\mathcal{R}} = \beta_2^{\mathcal{E}} = -1$. The target estimand is still constant, with $\tau = 0$ used to evaluate type I error and $\tau=0.4$ to evaluate power. This simulation setting also satisfies Assumption \ref{assum:exchange_delta} (i).

True values of the input parameters and models for the pre-experimental variance estimation process were: $\sigma_{0,0}^2 = 2.1$, $\sigma_{0,0}^2(\boldsymbol{X}) = 0.16X_1^4$, $\kappa_0^2 = \kappa_1^2 = 1.024$, $r(\boldsymbol{X})=3.2/X_1^2$, $r_0^M=r_1^M = 1.524/2.1$, $\gamma_1 = 1$, $\gamma = 1$, $d(\boldsymbol{X}) = 1$.


\subsubsection{Comparison of sample sizes}\label{appendix:sensi_3}

For each sensitivity analysis scenario, we calculated the true required sample size for each design method ($N_{\mathcal{R}}^{std,true}$, $N_{\mathcal{R}}^{aipw,true}$, $N_{\mathcal{R}}^{ec,true}$, $N_{\mathcal{R}}^{sa,true}$) using the true values of input parameters and models in the pre-experimental variance estimation procedure. We also computed the non-informative required sample size for the hybrid and single-arm design methods ($N_{\mathcal{R}}^{ec}$, $N_{\mathcal{R}}^{sa}$) based on the default settings through the variance estimation procedure. For each non-informative required sample size, we also conducted 2000 replications and took the average computed sample size to account for the variation due to the differences caused by estimating $\sigma_{0,0}^2$ and $\sigma_{0,0}^2(\boldsymbol{X})$ using different EC data.

\begin{table}[ht]
\centering
\small
\caption{\textit{Required sample sizes for different design methods under sensitivity analysis simulation settings. The methods include the RCT design with the difference-in-means estimator ($N_{\mathcal{R}}^{std,true}$), the RCT design using the AIPW estimator ($N_{\mathcal{R}}^{aipw,true}$), the hybrid design ($N_{\mathcal{R}}^{ec,true}$, $N_{\mathcal{R}}^{ec}$), and the single-arm design ($N_{\mathcal{R}}^{sa,true}$, $N_{\mathcal{R}}^{sa}$). The percentage of the required sample size saved relative to the RCT using the difference-in-means estimator, are denoted by $R^{aipw,true}$, $R^{ec,true}$ and $R^{sa,true}$, corresponding to the AIPW estimator, hybrid design, and single-arm design, respectively.}}
\begin{tabular}{cccccccccc} 
\hline
 \hline
 \multicolumn{10}{c}{\textbf{Heterogeneous Treatment Effect Across the Current Study Population ($N_{\mathcal{E}} = 1000$)}}\\
 \hline 
   $\pi_A$&  $N_{\mathcal{R}}^{std,true}$ & $N_{\mathcal{R}}^{aipw,true}$  & $N_{\mathcal{R}}^{ec, true}$ &
 $N_{\mathcal{R}}^{ec}$ & $N_{\mathcal{R}}^{sa,true}$ &$N_{\mathcal{R}}^{sa}$ & $R^{aipw, true}$ & $R^{ec, true}$& $R^{sa, true}$ \\
 \hline
0.5 & 286 & 162 & 88 & 103 & 47& 52 & 43.36\% & 69.23\% & 83.57\% \\
0.6 & 292 & 169 & 74 & 86 & 47 & 52 & 42.12\% & 74.66\% & 83.90\%\\
0.7 & 326 & 192 & 65 & 74 & 47& 52 & 41.10\% & 80.06\% & 85.58\% \\
0.8 & 418 & 251 & 57 & 65 & 47 & 52 & 39.95\% & 86.36\%  & 88.76\% \\
0.9 & 726 & 441 & 51 & 58 & 47 & 52 & 39.26\% & 92.98\% & 93.53\%\\
\hline
\hline
\multicolumn{10}{c}{\textbf{Non-constant Conditional Variance of the Outcome ($N_{\mathcal{E}} = 1000$)}}\\
 \hline 
 $\pi_A$& 
   $N_{\mathcal{R}}^{std,true}$ & $N_{\mathcal{R}}^{aipw,true}$  & $N_{\mathcal{R}}^{ec,true}$  & $N_{\mathcal{R}}^{ec}$ & $N_{\mathcal{R}}^{sa,true}$ & 
   $N_{\mathcal{R}}^{sa}$ &
   $R^{aipw,true}$ &
   $R^{ec,true}$ &
   $R^{sa,true}$ \\
 \hline
0.5 & 300 & 201 & 109 & 170 & 55 & 86
&33.00\% & 63.67\% & 81.67\% \\
0.6 & 312 & 210 & 91  & 142 & 55 &  86 &32.69\% & 70.83\% & 82.37\% \\
0.7 & 357 & 240 & 78  & 122 & 55 & 86
& 32.77\% & 78.15\%& 84.59\% \\
0.8 & 468 & 314 & 69  & 107 & 55 & 86 
& 32.91\% & 85.26\% & 88.25\% \\
0.9 & 832 & 557 & 61  & 95 & 55 & 86 &33.05\% & 92.67\% & 93.39\%\\
\hline
\hline
\end{tabular}
\label{app:table_sensi}
\end{table}

Table \ref{app:table_sensi} shows that, across all values of $\pi_A$, the true required sample sizes for the hybrid and single-arm design methods are consistently smaller than those for the RCT design methods using either the difference-in-means estimator or the AIPW estimator. Moreover, as $\pi_A$ increases, the percentage of sample size saved by both the hybrid design and the single-arm design methods increases accordingly.

In the heterogeneous treatment effect setting, compared to the RCT design with the difference-in-means estimator, the RCT design using the AIPW estimator achieves a sample size reduction of 39.26\% to 43.36\%. In contrast, the hybrid and single-arm design methods achieve much greater reductions, over 69.23\% and 83.57\%, respectively. In the setting with non-constant conditional variance of the outcome, the RCT design with the AIPW estimator achieves a reduction of 32.69\% to 33.00\%, while the hybrid design saves over 63.67\%, and the single-arm design achieves reductions exceeding 81.67\% across all $\pi_A$ values.

\subsubsection{Assessment performance in trial designs}\label{appendix:sensi_4}

Next, we assessed the performance of each estimator through the hypothesis test: $H_0: \tau = 0\ vs\ H_a: \tau\neq 0$. For each $\pi_A$, we varied $N_{\mathcal{R}}$ from $6$ to $600$, repeating the experiment $2000$ times per setting. We computed power (when $\tau = 0.4$) or type I error (when $\tau = 0$) for each estimator by determining the proportion of significant results in each scenario.


\begin{figure}[H]
\centering
\includegraphics[width=0.99\linewidth]{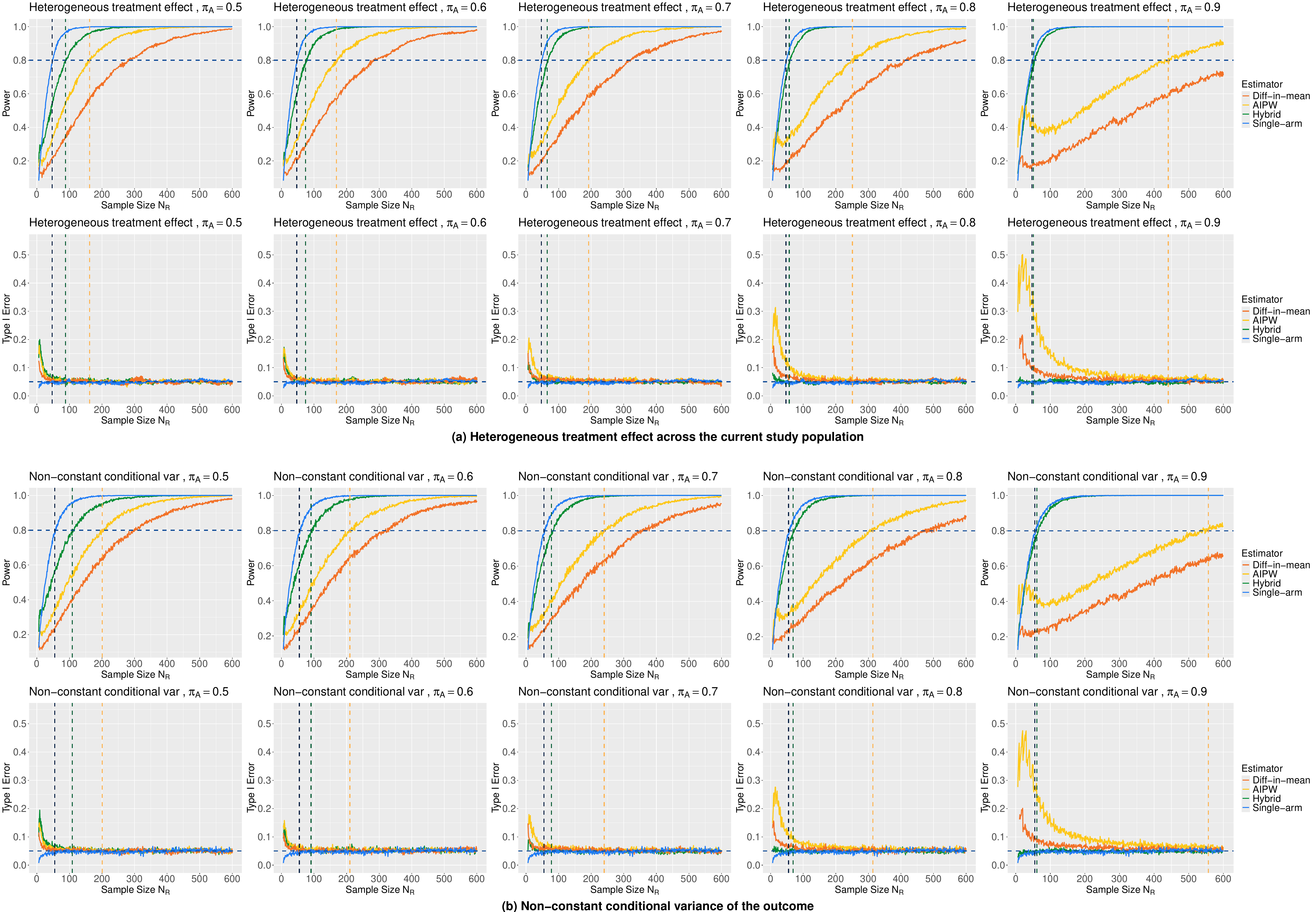}
\caption{Power and type I error of different design methods in sensitivity analysis simulation settings with sufficient EC data 
($N_{\mathcal{E}} = 1000$) under $\pi_A = 0.5,0.6,0.7,0.8, 0.9$. The design methods include the RCT design using the difference-in-means estimator, the RCT design using the AIPW estimator, the hybrid design method, and the single-arm design method. In each subfigure, the upper panel displays the empirical power, and the lower panel shows the type I error. The horizontal dashed lines represent the target power level $0.8$ in the power plot and the target type I error $0.05$ in the type I error plot. Vertical dashed lines indicate the true required sample sizes for each design method, shown from left to right: blue for $N_{\mathcal{R}}^{sa,true}$, green for $N_{\mathcal{R}}^{ec,true}$, and orange for $N_{\mathcal{R}}^{aipw,true}$. The curves represent the empirical power and the type I error across varying sample size $N_{\mathcal{R}}$.} 
\label{appendix:fig_sensi}
\end{figure}


In the setting with non-constant conditional variance of the outcome, although the true model for $r(\boldsymbol{X})$ is not constant, we estimated it using the empirical variance ratio estimator proposed by \cite{gao2025improving}: $\widehat{r}(\boldsymbol{X}) = N_{c}^{-2}\sum_{i\in\mathcal{R}} (1-A_i)\{Y_i-\widehat{\mu}_0(\boldsymbol{X}_i)\}^2 / N_{\mathcal{E}}^{-1}\sum_{i\in\mathcal{E}}\{Y_i-\widehat{\mu}_0(\boldsymbol{X}_i)\}^2$. Simulation results in Figure \ref{appendix:fig_sensi} indicate that both the hybrid and single-arm designs achieve power exceeding $0.8$ and maintain type I error close to the nominal level of $0.05$ at their true required sample sizes $N_{\mathcal{R}}^{ec,true}$ and $N_{\mathcal{R}}^{sa,true}$, across all sensitivity analysis settings. In addition, the hybrid design using $\widehat{\tau}_{ec}$ and the single-arm design using $\widehat{\tau}_{sa}$ have higher empirical power than the RCT designs using either the AIPW estimator or the difference-in-means estimator.

\end{document}